\documentclass[11pt]{article}

\usepackage{epsfig}
\usepackage{latexsym}
\usepackage{graphicx}
\usepackage{color}
\usepackage{amsmath,amssymb}
\usepackage{cite}

\makeatletter
\def\section{\@startsection {section}{1}{\z@}{-3.5ex plus -1ex minus -.2ex}{2.3ex plus .2ex}{\large\bf}}
\def\subsection{\@startsection{subsection}{2}{\z@}{-3.25ex plus -1ex
minus -.2ex}{1.5ex plus .2ex}{\normalsize\bf}}
\makeatother
\newcommand{\captionfonts}{\small}
\makeatletter  
\long\def\@makecaption#1#2{%
  \vskip\abovecaptionskip
  \sbox\@tempboxa{{\captionfonts #1: #2}}%
  \ifdim \wd\@tempboxa >\hsize
    {\captionfonts #1: #2\par}
  \else
    \hbox to\hsize{\hfil\box\@tempboxa\hfil}%
  \fi
  \vskip\belowcaptionskip}
\makeatother   

\catcode`@=11
\def\marginnote#1{}
\newcount\hour
\newcount\minute
\newtoks\amorpm
\hour=\time\divide\hour
by60
\minute=\time{\multiply\hour by60 \global\advance\minute
by-\hour}
\edef\standardtime{{\ifnum\hour<12 \global\amorpm={am}
\else\global\amorpm={pm}\advance\hour by-12 \fi
 \ifnum\hour=0
\hour=12 \fi
 \number\hour:\ifnum\minute<10
0\fi\number\minute\the\amorpm}}
\edef\militarytime{\number\hour:\ifnum\minute<10
0\fi\number\minute}
\def\draftlabel#1{{\@bsphack\if@filesw
{\let\thepage\relax
 \xdef\@gtempa{\write\@auxout{\string
\newlabel{#1}{{\@currentlabel}{\thepage}}}}}\@gtempa
 \if@nobreak
\ifvmode\nobreak\fi\fi\fi\@esphack}
\gdef\@eqnlabel{#1}}
\def\@eqnlabel{}
\def\@vacuum{}
\def\draftmarginnote#1{\marginpar{\raggedright\scriptsize\tt#1}}
\def\draft{\oddsidemargin
0.0truein
 \def\@oddfoot{\sl preliminary draft \hfil
\rm\thepage\hfil\sl\today\quad\militarytime}
 \let\@evenfoot\@oddfoot
\overfullrule 3pt
 \let\label=\draftlabel
\let\marginnote=\draftmarginnote
\def\@eqnnum{(\theequation)\rlap{\kern\marginparsep\tt\@eqnlabel}
\global\let\@eqnlabel\@vacuum}
}
\catcode`@=12

\newcommand{\beq}{\begin{eqnarray}}
\newcommand{\eeq}{\end{eqnarray}}
\newcommand{\bpmatrix}{\begin{pmatrix}}
\newcommand{\epmatrix}{\end{pmatrix}}
\newcommand{\fr}{\frac}
\newcommand{\la}{\lambda}
\newcommand{\crn}{\nonumber \\}
\newcommand{\ep}{\epsilon}
\newcommand{\hc}{\text{ h.c}}
\newcommand{\appen}[1]{Appendix~\ref{#1}}
\newcommand{\diag}{\text{diag}}
\newcommand{\eff}{\text{eff}}

\newcommand{\DRb}{\overline{\text{DR}}}
\newcommand{\al}{\alpha}
\newcommand{\RE}{\operatorname{Re}}
\newcommand{\Ga}{\Gamma}
\newcommand{\ga}{\gamma}
\newcommand{\ba}{\begin{array}}
\newcommand{\ea}{\end{array}}
\newcommand{\Si}{\Sigma}
\newcommand{\MSb}{\overline{\text{MS}}}
\newcommand{\eg}{{\it e.g.\;}}
\newcommand{\be}{\begin{equation}}
\newcommand{\ee}{\end{equation}}
\newcommand{\braket}[1]{\left(#1\right)}
\newcommand{\hs}{\hspace*{3mm}}
\newcommand{\De}{\Delta}
\newcommand{\de}{\delta}
\newcommand{\ti}{\tilde}
\newcommand{\bfZ}{\textbf{Z}}
\newcommand{\eq}[1]{Eq.~(\ref{#1})}
\newcommand{\fig}[1]{Fig.~\ref{#1}}
\newcommand{\tree}{\text{tree}}
\newcommand{\gev}{~\text{GeV}}
\newcommand{\ie}{{\it i.e.\;}}
\newcommand{\ssect}[1]{Subsection~\ref{#1}}
\newcommand{\sect}[1]{Section~\ref{#1}}
\newcommand{\mev}{~\text{MeV}}
\newcommand{\tev}{~\text{TeV}}
\newcommand{\ben}{\begin{enumerate}}
\newcommand{\een}{\end{enumerate}}
\newcommand{\higgsbound}{{\texttt{HiggsBounds-3.8.1}}}
\newcommand{\favn}{{\texttt{FeynArts-3.6}}}
\newcommand{\fcvn}{{\texttt{FormCalc-6.1}}}
\newcommand{\bc}{\begin{center}}
\newcommand{\ec}{\end{center}}
\newcommand{\si}{\sigma}

\def\mchi01{m_{\tilde{\chi}^0_1}}
\def\mst1{m_{\tilde{t}_1}}
\def\msc1{m_{\tilde{c}_L}}
\def\mn1{m_{\tilde{\chi}_1^0}}

\def\acdj1{a_{j1}^{\tilde{c} d_k}}
\def\atdj1{a_{j1}^{\tilde{t} d_k}}
\def\bcdj1{b_{j1}^{\tilde{c} d_k}}
\def\btdj1{b_{j1}^{\tilde{t} d_k}}

\newcommand{\lsim}{\raisebox{-0.13cm}{~\shortstack{$<$ \\[-0.07cm]
      $\sim$}}~}
\newcommand{\s}{\newline \vspace*{-3.5mm}}
\newcommand{\id}{{\rm 1\kern-.12em
\rule{0.3pt}{1.5ex}\raisebox{0.0ex}{\rule{0.1em}{0.3pt}}}}

\begin{document}

\thispagestyle{empty}

\begin{center}
\rightline{KA-TP-14-2013}
\rightline{SFB/CPP-13-41}

\begin{center}

\vspace{1.7cm}

{\LARGE{\bf Higher Order Corrections to the Trilinear Higgs \\[0.1cm] 
Self-Couplings in the Real NMSSM}}
\end{center}

\vspace{1.4cm}

{\bf Dao Thi Nhung}, {\bf Margarete M\"uhlleitner}, {\bf
  Juraj Streicher} and {\bf Kathrin Walz} \\

\vspace{1.2cm}

{\em Institut f\"ur Theoretische Physik, Karlsruher Institut f\"ur  Technologie, \\
76128 Karlsruhe, Germany}

\end{center}

\vspace{0.8cm}

\centerline{\bf Abstract}
\vspace{2 mm}
\begin{quote}
\small
After the discovery of a Higgs-like boson by the LHC experiments ATLAS
and CMS, it is of crucial importance to determine its properties in
order to not only identify it as the boson responsible for electroweak
symmetry breaking but also to clarify the question if it is a Standard
Model (SM) Higgs boson or the Higgs particle of some extension beyond
the SM as {\it e.g.}~supersymmetry. In this context, the precise prediction of the Higgs
parameters as masses and couplings plays a crucial role for the 
proper distinction between different models. In extension of previous works
on the loop-corrected Higgs boson masses of the Next-to-Minimal
Supersymmetric Extension of the SM (NMSSM), we present here the
calculation of the loop-corrected trilinear NMSSM Higgs self-couplings.
The loop corrections turn out to have a substantial 
impact on the decay widths of Higgs-to-Higgs decays and on the production
cross section of Higgs pairs via gluon fusion. They are therefore
indispensable for the correct interpretation of the experimental Higgs
results.
\end{quote}

\newpage


\section{Introduction}
In 2012 the Large Hadron Collider (LHC) experiments ATLAS and CMS
announced the discovery of a new scalar particle \cite{:2012gk,:2012gu}. Since then
experimental and theoretical activities have started to pin 
down the true nature of this particle. It has to be clarified if the particle is
really the Higgs boson, {\it i.e.}~the particle responsible for 
electroweak symmetry breaking (EWSB) without violating the gauge
principles of the Standard Model (SM). And if so, whether it is the Higgs
boson of the SM or one of an 
enlarged supersymmetric (SUSY) Higgs sector or some more exotic
version of the Higgs particle, like {\it e.g.} a composite object. To
this aim, the coupling strengths of the new particle to the other SM
particles, its spin and CP-properties and finally its trilinear and
quartic self-couplings have to be measured. While the absolute coupling values
are not accessible at the LHC, fits can be performed to the measured
signal strengths in the various Higgs search channels \cite{higgsfits}. 
The Higgs spin and CP quantum numbers can be extracted
from angular and threshold distributions in various Higgs production
and decay channels \cite{higgsspin}. The trilinear and quartic Higgs self-interactions
finally are in principle accessible in double and triple Higgs
production \cite{Djouadi:1999gv,Djouadi:1999rca,tripleh,Baglio:2012np,quartic}. The
knowledge of these couplings enables the reconstruction of the Higgs
potential and allows to test if it has a non-vanishing vacuum
expectation value (VEV) as required by the Higgs mechanism. This
challenging experimental program necessitates on the theoretical side
the precise prediction of production and decay cross sections in the model under
consideration, in order to be able to interpret the experimental data
correctly and to distinguish between different models. The cross sections
have therefore to be evaluated including higher order
corrections. However, not only these, but
also the input parameters like masses and couplings have to be
predicted with highest possible precision. It is well known {\it e.g.}~that in
the Minimal Supersymmetric extension of the SM (MSSM) the lightest
Higgs boson mass is shifted beyond the theoretical tree-level bound of the $Z$
boson mass only through the inclusion of higher-order corrections
\cite{Djouadi:2005gj}. In this work we contribute to increasing the
accuracy in the prediction of the Higgs parameters of the
Next-to-Minimal Supersymmetric extension of the SM (NMSSM)
\cite{genNMSSM1,genNMSSM2,Nevzorov:2004ge}. We
calculate the one-loop corrections to the trilinear Higgs
self-couplings of the NMSSM Higgs bosons in the Feynman-diagrammatic
approach, after having provided in previous works the one-loop 
corrections to the masses \cite{Ender:2011qh,Graf:2012hh}. \s 

The Higgs sector of the NMSSM consists of two complex Higgs doublets
$H_u$ and $H_d$ and an additional complex singlet field $S$. The
singlet field couples to the MSSM Higgs doublets through the interaction
term $\lambda S (H_u \epsilon H_d)$. This allows for a dynamical
solution of the $\mu$ problem \cite{MuProblem} when the neutral component of
the singlet field acquires its VEV. Moreover, the NMSSM requires less
fine-tuning than the MSSM in order to comply with the LHC discovery of a Higgs boson
with mass around 125~GeV \cite{finetune}. New contributions
proportional to the quartic coupling 
increase the tree-level mass value of the lightest Higgs boson, so that
compared to the MSSM less important radiative mass corrections are
necessary to shift the tree-level mass value to the observed
125~GeV. This in turn allows for lighter stop masses and hence less
fine-tuning. After EWSB we are left with seven Higgs bosons, which are
in the CP-conserving NMSSM three neutral CP-even, two
neutral CP-odd and two charged Higgs bosons. The enlarged Higgs sector
leads to interesting phenomenological implications. Thus the SM-like
Higgs boson, which is compatible with the LHC Higgs searches, can in
general be either of the three neutral CP-even Higgs
bosons. Most scenarios, however, which are in accordance with the
experimental constraints, feature the lightest or the next-to-lightest
CP-even Higgs boson as the SM-like 125~GeV boson. Furthermore, the
admixture of the singlet field  
can suppress the Higgs couplings to the other SM particles, so that
light Higgs states may have escaped detection at Tevatron, LEP 
and LHC. The presence of light Higgs bosons entails possible new
Higgs-to-Higgs decays such as {\it e.g.} the decay of a SM-like scalar
Higgs boson into a pair of lighter pseudoscalars. From this discussion
it becomes clear that the precise knowledge of the Higgs boson masses and
couplings is inevitable to properly describe the Higgs phenomenology.
It is needed to reanalyse and interpret correctly the LHC search results in
the light of a possible NMSSM extension of the SM. \s

While in the MSSM the higher-order corrections to the Higgs boson
masses have been calculated up to the inclusion of the leading
contributions from three-loop order \cite{Djouadi:2005gj}, the
higher-order corrections to 
the NMSSM Higgs boson masses have not reached the same level of
accuracy. For the CP conserving NMSSM the following corrections are
available. In the effective potential approach the leading one-loop (s)top and
(s)bottom contributions have been calculated \cite{effpot}. The chargino,
neutralino and scalar one-loop contributions are available at leading
logarithmic accuracy \cite{leadlog}. The full one-loop
contributions have been given in the $\overline{\mbox{DR}}$
renormalisation scheme in Ref.~\cite{Degrassi:2009yq,full1loop}, the ${\cal O}
(\alpha_t \alpha_s + \alpha_b \alpha_s)$ corrections in the
approximation of zero external momentum in Ref.~\cite{Degrassi:2009yq}. In
addition, we have provided the full one-loop corrections 
in a mixed $\overline{\mbox{DR}}$-on-shell and in a pure on-shell
renormalisation scheme \cite{Ender:2011qh}. As for the CP-violating NMSSM,
CP-violating effects from the third generation squark sector, from the
charged particle loops and from gauge boson contributions have been
provided in the effective potential approach at one-loop level
\cite{effcorr1,effcorr2,effcorr3}. The full one-loop and
logarithmically enhanced two-loop 
effects are available in the renormalisation group improved approach
\cite{complex2loop}. This has been complemented by the full one-loop
corrections in the Feynman diagrammatic approach \cite{Graf:2012hh}. \s

Both Higgs boson masses and Higgs self-interactions arise from the
Higgs potential. They cannot be separated from each other. In order to
consistently describe the Higgs sector including higher-order
corrections, it is therefore not sufficient to only correct the Higgs
boson masses. The trilinear and quartic Higgs self-interactions
have to be evaluated at the same order in perturbation theory and
within the same renormalisation scheme as the Higgs boson masses 
to allow for a consistent description of the Higgs boson
phenomenology. While the phenomenology involving quartic Higgs
self-couplings is most probably outside the reach of existing and
future colliders, the trilinear Higgs self-couplings
play a role in the determination of the Higgs boson branching
ratios into SM particles, in the evaluation of Higgs-to-Higgs decays
and in Higgs pair production processes. In this work we calculate the one-loop
corrections to the trilinear Higgs self-couplings of the CP-conserving
NMSSM in the Feynman-diagrammatic approach. We apply the mixed
$\overline{\mbox{DR}}$-on-shell renormalisation scheme, which has been 
introduced in Ref.~\cite{Ender:2011qh} for the one-loop corrections to the Higgs
boson masses, and we study the phenomenological implications of these
corrections. \s

The outline of our paper is as follows. In section
\ref{sec:HiggsMassNMSSM} we briefly describe the loop corrections to
the Higgs boson masses. We use this section to set up our notation,
present details of the calculation of the Higgs mass corrections and
introduce the renormalisation scheme. In contrast to our previous
calculation \cite{Ender:2011qh} we also add leading two-loop
contributions which have been taken over from
Ref.~\cite{Degrassi:2009yq}. Section \ref{sec:higgstohiggs} contains the calculation
of the loop-corrected trilinear Higgs self-couplings. Section
\ref{sect-results} is devoted to our numerical analysis. We first
define in subsection \ref{sect-constraints} our input parameters and
describe the constraints which we apply. In particular we seek a
Higgs boson with mass around 125~GeV that is compatible with the LHC
results for the signal strengths in the various final states, while making sure that the
remaining Higgs mass spectrum has not been excluded yet. In subsection
\ref{sec:effective} we discuss the effective trilinear Higgs couplings before we
present in subsection \ref{sec:resdecays} our results on Higgs-to-Higgs
decays. Subsection \ref{sec:pairprod} is devoted to the discussion of the effects
of loop corrections to the trilinear Higgs self-couplings on Higgs pair
production processes at the LHC. We conclude in section \ref{sec:conclusions}.
In the Appendix \ref{app:1} we list the tree-level trilinear
Higgs couplings.

\section{The loop-corrected Higgs boson masses}
\label{sec:HiggsMassNMSSM}
In this section we summarise the calculation of the loop corrections
to the NMSSM Higgs boson masses. Since at one-loop order we apply the same
procedure as in our 
previous publication \cite{Ender:2011qh} we repeat here only the main
features for the purpose of setting up the notation and 
of introducing the renormalisation scheme. For details we refer the
reader to Ref.~\cite{Ender:2011qh}. \s 
 
We work in the framework of the NMSSM with a scale
invariant superpotential. The Higgs mass matrix is derived from the
corresponding NMSSM Higgs potential, which is obtained from the
superpotential $W_{NMSSM}$ of the NMSSM, the soft SUSY breaking terms
and the $D$-term contributions. In terms of the Higgs doublet
superfields $\hat{H}_u$ and $\hat{H}_d$, which couple to the up- and down-type
fermion superfields, respectively, and of the singlet superfield
$\hat{S}$, the superpotential is given by
\beq
W_{NMSSM} = W_{MSSM} - \epsilon_{ij} \lambda \hat{S} \hat{H}^i_d
\hat{H}^j_u + \frac{1}{3} \kappa \hat{S}^3 \;.
\eeq
The indices of the $SU(2)_L$ fundamental representation are denoted by
$i,j=1,2$, and $\epsilon_{ij}$ is the totally antisymmetric tensor
with $\epsilon_{12}= \epsilon^{12} = 1$. The dimensionless parameters
$\lambda$ and $\kappa$ are taken to be real as we assume 
CP conservation. The MSSM superpotential $W_{MSSM}$ reads in terms of the
quark and lepton superfields and their charge conjugate (denoted by the
superscript $c$), $\hat{Q}, \hat{U}^c, \hat{D}^c, \hat{L},
\hat{E}^c$, 
\beq
W_{MSSM} = \epsilon_{ij} [y_e \hat{H}^i_d \hat{L}^j \hat{E}^c + y_d
\hat{H}_d^i \hat{Q}^j \hat{D}^c - y_u \hat{H}_u^i \hat{Q}^j \hat{U}^c] \;.
\label{eq:superpotmssm}
\eeq
For simplicity colour and generation indices have been
suppressed. Following common NMSSM constructions we have assumed the
MSSM $\mu$ term to be zero and also terms linear and quadratic in
$\hat{S}$. The NMSSM soft SUSY breaking Lagrangian expressed in terms of the
component fields $H_u, H_d$ and $S$ reads
\beq
\mathcal L_{soft} = {\cal L}_{soft,\, MSSM} - m_S^2 |S|^2 +
(\epsilon_{ij} \lambda 
A_\lambda S H_d^i H_u^j - \frac{1}{3} \kappa
A_\kappa S^3 + h.c.) \;. 
\eeq
It contains the soft SUSY breaking MSSM contribution
\beq
{\cal L}_{soft, \, MSSM} &=& -m_{H_d}^2 H_d^\dagger H_d - m_{H_u}^2
H_u^\dagger H_u -
m_Q^2 \tilde{Q}^\dagger \tilde{Q} - m_L^2 \tilde{L}^\dagger \tilde{L}
- m_U^2 \tilde{u}_R^* 
\tilde{u}_R - m_D^2 \tilde{d}_R^* \tilde{d}_R 
\nonumber \\\nonumber
&& - m_E^2 \tilde{e}_R^* \tilde{e}_R - (\epsilon_{ij} [y_e A_e H_d^i
\tilde{L}^j \tilde{e}_R^* + y_d
A_d H_d^i \tilde{Q}^j \tilde{d}_R^* - y_u A_u H_u^i \tilde{Q}^j
\tilde{u}_R^*] + h.c.) \\
&& -\frac{1}{2}(M_1 \tilde{B}\tilde{B} + M_2
\tilde{W}_k\tilde{W}_k + M_3 \tilde{G}\tilde{G} + h.c.) \;,
\label{eq:softmssm}
\eeq
where $\tilde{Q}=(\tilde{u}_L,\tilde{d}_L)^T$ and
$\tilde{L}=(\tilde{\nu}_L,\tilde{e}_L)^T$ with tilde denoting the scalar
components of the corresponding quark and lepton superfields. The soft
SUSY breaking gaugino mass terms for the gaugino fields $\tilde{B},
\tilde{W}_k$ ($k=1,2,3$) and $\tilde{G}$ are summarised in the last
line of Eq.~(\ref{eq:softmssm}) (the summation over paired indices is
implicit). Working in the CP-invariant NMSSM the 
soft SUSY breaking trilinear couplings $A_x$
($x=\lambda,\kappa,d,u,e$) and gaugino mass parameters $M_k$ are
taken to be real. Furthermore, squark and slepton mixing between the
generations is neglected. Like in the majority of phenomenological NMSSM
constructions we have omitted possible soft SUSY breaking terms linear
and quadratic in the singlet field $S$. After electroweak symmetry
breaking the neutral components of the Higgs doublet and singlet
fields acquire non-vanishing vacuum expectation values. Substituting
the expansions of the Higgs fields about their 
VEVs $v_u, v_d$ and $v_s$, which are chosen to be real and positive, 
\beq
H_d =
 \bpmatrix (v_d + h_d +ia_d)/{\sqrt 2}\\ h_d^- \epmatrix,\quad
H_u = \bpmatrix
h_u^+ \\ (v_u + h_u +ia_u)/{\sqrt 2}\epmatrix,\quad
S= \fr{v_s + h_s +ia_s}{\sqrt 2} \;,
\label{eq:Higgs_decomposition} 
\eeq
into the Higgs potential $V_H$, which expressed in terms of the Higgs component
fields reads,
\beq
V_{H}  &=& (|\la S|^2 + m_{H_d}^2)H_{d,i}^* H_{d,i}+ (|\la S|^2 + m_{H_u}^2)H_{u,i}^* H_{u,i} +m_S^2 |S|^2 \crn
&& + \fr18 (g_2^2+g_1^{2})(H_{d,i}^* H_{d,i}-H_{u,i}^* H_{u,i} )^2
+\fr12g_2^2|H_{d,i}^* H_{u,i}|^2\crn
&&   + |-\ep^{ij} \la  H_{d,i}  H_{u,j} + \kappa S^2 |^2+
\big[-\ep^{ij}\la A_\la S   H_{d,i}  H_{u,j}  +\fr13 \kappa
A_{\kappa} S^3+\hc \big] \,,
\label{eq:higgs_potential} 
\eeq
we have 
\begin{align} V_H = V_H^{\mbox{\scriptsize const}} & +  t_{h_d} h_d + t_{h_u} h_u +  t_{h_s} h_s  
+ 
 \fr 12 \bpmatrix h_d,h_u,h_s \epmatrix  {M_S^2}
\bpmatrix  h_d\\h_u\\h_s \epmatrix + \fr 12 \bpmatrix
a_d,a_u,a_s\epmatrix {M_a^2} 
\bpmatrix a_d\\a_u\\a_s \epmatrix
\crn &+
  \bpmatrix h_d^-,h_u^-\epmatrix {M_C^2}
\bpmatrix  h_d^+\\ h_u^+\epmatrix  +\la^{hhh}_{ijk}h_ih_jh_k+ \la^{haa}_{ijk}h_ia_ja_k+
V_H^{4\phi} \,,
\label{eq:vhpot}
\end{align}
with $i,j,k=d,u,s$. 
Equation~(\ref{eq:vhpot}) contains the tadpole coefficients $t_{h_i}$ of the
terms linear in the Higgs fields $h_i$ , the mass matrices squared and the
trilinear Higgs self-interactions. 
The constant terms are summarised in 
$V_H^{\mbox{\scriptsize const}}$ and the quartic Higgs interactions in
$V_H^{4\phi}$. They are not specified here as they are not needed for
our calculation. The $3\times 3$ mass matrices squared for the neutral
CP-even and CP-odd Higgs sector, respectively, are denoted by
${M_S^2}$ and ${M_a^2}$, the $2\times 2$ charged Higgs mass
matrix squared by 
${M_C^2}$ and the trilinear Higgs self-couplings by
$\lambda_{ijk}^{\phi \phi' \phi''}$. Explicit expressions for the
couplings are given in \appen{app:1}. Performing a first 
rotation of the CP-odd fields $(a_d, a_u, a_s)$,
\vspace*{0.3cm}
\beq
\left( \begin{array}{c} a \\ a_s \\ G\end{array} \right) = {\cal R}_G
\left( \begin{array}{c} a_d \\ a_u \\ a_s \end{array} \right) \equiv
\left( \begin{array}{ccc} s_{\beta_n} & c_{\beta_n} & 0 \\
 0 & 0 & 1 \\
 c_{\beta_n} & -s_{\beta_n} & 0 
\end{array} \right)
\left( \begin{array}{c} a_d \\ a_u \\ a_s \end{array} \right)  \;,
\label{eq:rotmatcpodd}
\eeq
allows to separate a massless Goldstone boson $G$ and yields the
pseudoscalar mass matrix squared
\beq
{M_P^2} = {\cal R}^G {M_a^2} ({\cal R}^{G})^T \;.
\eeq
Here and in the following we adopt the shorthand notation $c_x \equiv \cos x,
s_x \equiv \sin x$. At tree-level the angle $\beta_n$ coincides with
the angle $\beta$ defined by the ratio of the two VEVs $v_u$ and $v_d$,
$\tan\beta = v_u/v_d$. 
Explicit expressions for the scalar and pseudoscalar mass matrices
squared ${M_S^2}$ and ${M_P^2}$ as well as for the tadpole
parameters can be found in \cite{Ender:2011qh}, and for the charged
Higgs mass matrix $M_C$ in \cite{Graf:2012hh}. The CP-even Higgs mass
eigenstates $h_m$ ($m=1,2,3$) are obtained via the diagonalisation of the mass
mixing matrix squared ${M_S^2}$ by an orthogonal transformation,
\beq
 \bpmatrix h_1,h_2,h_3 \epmatrix^T =   ({\cal R}^{S}) \bpmatrix
 h_d,h_u,h_s \epmatrix^T, 
\quad \diag(m_{h_1}^2,m_{h_2}^2,m_{h_3}^2)= {\cal R}^S {M_S^2}
({\cal R}^S)^T \label{eq:rotationS} \; .
\eeq
The mass eigenstates are ordered by ascending mass with $m_{h_1} \le
m_{h_2} \le m_{h_3}$. The CP-odd mass eigenstates $a_i \equiv a_1,
a_2, G$ $(i=1,2,3)$ are obtained via an orthogonal rotation ${\cal R}^P$ applied
to $h^P=(a,a_s,G)^T$,  
\beq
a_i = {\cal R}^P_{ij} h_j^P \qquad i,j=1,2,3 \;,
\label{eq:rotationP}
\eeq
yielding the diagonal mass matrix squared
\beq
\diag(m_{a_1}^2,m_{a_2}^2,0) = {\cal R}^P \, {M_P^2} \, ({\cal R}^P)^T =
{\cal R}^P {\cal R}^G \, {M_a^2} \, ( {\cal R^P} {\cal R}^G )^T \;.
\eeq
Note that at tree-level ${\cal R}^P_{33}=1$ and ${\cal R}^P_{3i}=
{\cal R}^P_{i3}=0$ for $i\ne 3$. \s

At lowest order, the Higgs potential is determined by a set of twelve
parameters consisting of the electroweak gauge couplings $g_1$ and
$g_2$, the three VEVs, the soft SUSY breaking mass parameters of
the doublet and singlet Higgs fields and the new NMSSM related
parameters and soft SUSY breaking couplings, hence
\beq 
g_1,g_2,v_d,v_u,v_s,m_{H_d}^2,m_{H_u}^2,m_S^2,\la, \kappa,A_\la,
A_\kappa  \;. \label{eq:parset1}
\eeq
For physical interpretations it is convenient to replace some of these
parameters. Thus the minimisation conditions of the Higgs potential
$V$ can be exploited to trade $m_{H_d}^2,m_{H_u}^2$ and $m_S^2$ for the
tadpole parameters $t_{h_d},t_{h_u}$ and $t_{h_s}$. The charged Higgs boson
mass $M_{H^\pm}$ is used 
instead of the soft SUSY breaking coupling $A_\lambda$ and the
parameters $g_1,g_2, v_u$ and $v_d$ are replaced by the gauge boson
masses $M_W$ and $M_Z$, the electric charge $e$ and $\tan\beta$. We are then
left with the 'physical' parameter set
\beq
M_Z,M_W, M_{H^\pm},t_{h_d},t_{h_u},t_{h_s},e,\tan\beta,\la,
v_s,\kappa, A_\kappa \;. \label{eq:parset2}
\eeq
The tree-level relations between the two parameter sets
of Eq.~(\ref{eq:parset1}) and Eq.~(\ref{eq:parset2}) can be found
in \cite{Ender:2011qh}. Note that although the three tadpole parameters
vanish at the stable minimum of the potential, they have been kept for the purpose of
renormalisation at loop level. The terms linear in the Higgs
fields get loop contributions at higher order, and the tadpole
parameters are renormalised such that the conditions of a stable
minimum are fulfilled by the Higgs potential. \s

The calculation of the loop corrections to the Higgs boson masses and
decays leads to ultraviolet divergences which can be absorbed by the
renormalisation of the parameters entering the loop calculation. We
choose here the mixed renormalisation scheme proposed in
\cite{Ender:2011qh}. In this scheme part of the parameters are
renormalised in the on-shell (OS) scheme, the remaining ones via $\DRb$
conditions. We slightly abuse the term on-shell condition as we
apply it also for the renormalisation conditions of the tadpole
parameters. In detail we have,
\beq
\underbrace{M_Z,M_W,M_{H^\pm},t_{h_u},t_{h_d},t_{h_s},e}_{\mbox{on-shell
 scheme}}, 
\underbrace{\tan \beta, \lambda, v_s, \kappa, 
A_\kappa}_{\overline{\mbox{DR}} \mbox{ scheme}} \;.
\label{eq:mixedcond}
\eeq
The corresponding counterterms are given explicitly in
Ref.~\cite{Ender:2011qh}. The only difference consists in the electric
charge $e$ where we use the fine structure
constant at  the $Z$ boson mass $M_Z$, $\al=\al(M_Z)$, as an input in
order to make the results independent of $\ln m_f$ from the light
fermions, $f\ne t$. The counterterm $\delta Z_e$ of $e$ is therefore modified as
\cite{Denner:1991kt}  
\beq
\delta Z_e^{\alpha(M_Z)}&=&\delta Z_e^{\alpha(0)} - \fr{1}{2}\Delta\alpha(M_Z^2),\crn
\Delta\alpha(M_Z^2)&=&\fr{\partial \Sigma_T^{AA}}{\partial 
  k^2}\bigg\vert_{k^2=0}-\fr{\RE\Sigma_T^{AA}(M_Z^2)}{M_Z^2} \;, 
\eeq
where the transverse part of the photon self-energy $\Sigma_T^{AA}$
includes only the light fermion contributions.
For the Higgs field wave functions we also use the $\DRb$
scheme. \s

The loop-corrected Higgs masses squared are determined
numerically. They are given as the zeros of the determinant of the
two-point vertex functions for the scalars, $\hat{\Gamma}^S (p^2)$, and
the pseudoscalars, $\hat{\Gamma}^P (p^2)$, respectively. For the
scalar bosons it reads
\beq
\hspace*{-1cm} \hat{\Gamma}^S(p^2)  &=& i (p^2 - \hat{M}_S^2 (p^2) )\nonumber \\
  && \hspace*{-2cm} =i \left(\ba{ccc}
 p^2 - m_{h_1}^2 + \hat{\Si}_{h_1h_1}(p^2) &  \hat{\Si}_{h_1h_2}(p^2) & 
\hat{\Si}_{h_1 h_3}(p^2) \\
     \hat{\Si}_{h_2h_1}(p^1) &  p^2 - m_{h_2}^2 + \hat{\Si}_{h_2 h_2}(p^2) &  
\hat{\Si}_{h_2 h_3}(p^2) \\
   \hat{\Si}_{h_3h_1}(p^2) & \hat{\Si}_{h_3 h_2}(p^2) &
   p^2 - m_{h_3}^2 +\hat{\Si}_{h_3 h_3}(p^2)
  \ea\right)\label{eq:mass_Higgs_matrix}
\eeq
and for the pseudoscalar Higgs bosons
\beq
\hat{\Gamma}^P(p^2)& = i (p^2 - \hat{M}_A^2 (p^2) ) =i \bpmatrix p^2 -m_{a_1}^2 + \hat{\Si}_{a_1 a_1}(p^2) & \hat{\Si}_{a_1 a_2}(p^2)\\
 \hat{\Si}_{a_2 a_1}(p^2) &  p^2 - m_{a_2}^2 +  \hat{\Si}_{a_2 a_2}(p^2)
 \epmatrix \;.
\label{eq:massmatrix} 
\eeq
By $\hat{\Sigma}$ we denote the renormalised self-energy built
from the unrenormalised self-energy and tadpole contributions evaluated
at one-loop order and the counterterms in the mixed renormalisation
scheme. They are functions of the external momentum squared $p^2$. We
furthermore included two-loop corrections, as will be 
explained below. The masses squared $m_{h_i}^2, m_{a_l}^2$ ($i=1,2,3$,
$l=1,2$) are taken at tree-level as
indicated by the small letter $m$ for the mass. Note that we have not taken into
account the mixing of the pseudoscalars $a_1$ and $a_2$ with the
Goldstone boson $G$ and the longitudinal component of the $Z$ boson,
as we have checked explicitly that this
mixing gives negligible contributions to the one-loop corrected Higgs
boson masses. However, the mixing is taken into account in the
computation of the loop corrections to the Higgs boson decays into
two lighter Higgs bosons. The mass eigenvalues
are obtained in an iterative procedure. For example, in order to get the
lightest CP-even Higgs mass, in the first step the external momentum
squared $p^2$ is set equal to the $h_1$ tree-level mass. The mass
matrix squared is then diagonalised and the thus obtained lightest mass
eigenvalue is reinserted into the self-energies, to calculate the mass
eigenvalue in the next iteration. This procedure is repeated until the
difference in the mass eigenvalue of two subsequent iterations is less
than $10^{-10}$~GeV. The eigenvalues are in general complex, and the loop
corrected Higgs boson masses are given by the real part. They are
sorted by ascending mass, as
\beq
M_{H_1}^2 <M_{H_2}^2 < M_{H_3}^2 \quad \mbox{and} \quad M_{A_1}^2 <
M_{A_2}^2 \;.
\eeq
We denote the loop-corrected masses by capital letters, $M_{H_i}$,
$M_{A_l}$ ($i=1,2,3$, $l=1,2$) contrary to the tree-level masses with
small letters.\s

As indicated above, in order to improve the loop-corrected Higgs
boson masses, we have included in the mass matrices $\hat{M}_S^2(p^2)$
and $\hat{M}_P^2(p^2)$ the known two-loop ${\cal
  O}(\al_s\al_t+\al_s\al_b)$ corrections \cite{Degrassi:2009yq}, which
have been evaluated in the limit of zero external momentum. In this
calculation the $\DRb$ renormalisation scheme has been applied in the
(s)top and (s)bottom sectors. Therefore, in order to use these
corrections consistently, we use the running $\DRb$ top and bottom
quark masses in the evaluation of the one-loop renormalised
self-energies. 
Using as input the top quark pole mass $M_t$, we first translate it to
the running $\MSb$ top mass $m^{\MSb}_t(M_t)$ by applying the two-loop
relation, see \eg \cite{Melnikov:2000qh} and references therein, 
\be 
m^{\MSb}_t(M_t)=\left( 1-\fr43\braket{\fr{\al_s(M_t)}{\pi}} -
9.1253\braket{\fr{\al_s(M_t)}{\pi}}^2 \right) M_t \; ,
\ee
where $\alpha_s$ is the strong coupling constant evaluated at two-loop
order. As for the bottom quark mass input, it is already given as an
$\MSb$ mass at the scale $m_b^{\MSb}$. Both $m^{\MSb}_t(M_t)$ and
$m^{\MSb}_b(m_b^{\MSb})$ are then evolved up to the renormalisation
scale $\mu_R$, by using the two-loop formula 
\begin{align}
m_t^{\MSb}(\mu_R)&=U_6(\mu_R, M_t)m^{\MSb}_t(M_t)  \hs \;
\mbox{for} \; \mu_R > M_t  
\; , \\[0.1cm]
 m_b^{\MSb}(\mu_R)&=
\left\{ \begin{array}{rcl}
U_6(\mu_R, M_t) \, U_5(M_t, m^{\MSb}_b) \, m^{\MSb}_b(m_b^{\MSb})\hs
\text{for}\hs \mu_R > M_t \; , \\[0.1cm]
U_5(\mu_R, m^{\MSb}_b) \, m^{\MSb}_b(m_b^{\MSb}) \hs \text{for}\hs
\mu_R \le M_t \; ,
\end{array}\right.
\label{mb_evolution}
\end{align}   
where the evolution factor $U_n$ reads (see \eg\ \cite{Carena:1999py})
\beq
U_n(Q_2,Q_1)&=&\left(\fr{\alpha_s(Q_2)}{\alpha_s(Q_1)}\right)^{d_n}\left[1 + 
\fr{\alpha_s(Q_1) - \alpha_s(Q_2)}{4\pi}J_n\right] \; ,\hs Q_2 > Q_1\crn
d_n&=&\fr{12}{33-2n} \; , \hs J_n = -\fr{8982 - 504n + 40n^2}{3(33 -
  2n)^2} \; ,
\eeq
with $n$ being the number of active quark flavors 
($n=5$ for $m^{\MSb}_b(m_b^{\MSb})<Q\le M_t$ and 
$n=6$ for $Q> M_t$). From the $\MSb$ masses the $\DRb$ masses are 
computed by using the two-loop relation \cite{Avdeev:1997sz},
\beq
m_{t/b}^{\DRb}(\mu_R)=m_{t/b}^{\MSb}(\mu_R)\left[1 -
  \fr{\alpha_s(\mu_R)}{3\pi} - \fr{\alpha_s^2(\mu_R)}{144\pi^2}(73-3n)
\right] \; .
\eeq   
In addition, the possibly large supersymmetric corrections are resummed into
the effective top and bottom quark masses
\cite{Carena:1999py,Pierce:1996zz,Carena:2002bb,Guasch:2003cv}, 
\be
m_t^\eff = \fr{m_t^{\DRb}(\mu_R)}{1+\de m_t +\De m_t/ \tan\beta} \quad
\mbox{and} \quad m_b^\eff = \fr{m_b^{\DRb}(\mu_R)}{1+\de m_b +\De m_b
  \tan\beta} \; ,
\ee
where
\begin{align}
\De m_t&= \fr{2\al_s(\mu_R)}{3\pi} \, m_{\ti g} \, \mu_{\eff} \,
I(m_{\ti t_1}^2,m_{\ti t_2}^2, m_{\ti g}^2) + \fr{y_b^2}{16\pi^2} \,
A_b \, \mu_\eff  \, I(m_{\ti b_1}^2,m_{\ti b_2}^2, \mu_\eff^2) \; ,\\
\De m_b&= \fr{2\al_s(\mu_R)}{3\pi} \, m_{\ti g} \, \mu_{\eff} \,
I(m_{\ti b_1}^2,m_{\ti b_2}^2, m_{\ti g}^2) + \fr{y_t^2}{16\pi^2} \,
A_t \, \mu_\eff \, I(m_{\ti t_1}^2,m_{\ti t_2}^2, \mu_\eff^2) \; ,\\
\de m_t&= -\fr{2\al_s(\mu_R)}{3\pi} \, m_{\ti g} \, A_t \, I(m_{\ti
  t_1}^2,m_{\ti t_2}^2, m_{\ti g}^2) - \fr{y_b^2}{16\pi^2} \,
\mu_\eff^2 \, I(m_{\ti b_1}^2,m_{\ti b_2}^2, \mu_\eff^2) \; ,\\
\de m_b&= -\fr{2\al_s(\mu_R)}{3\pi} \, m_{\ti g} \, A_b I(m_{\ti
  b_1}^2,m_{\ti b_2}^2, m_{\ti g}^2) - \fr{y_t^2}{16\pi^2} \,
\mu_\eff^2 \, I(m_{\ti t_1}^2,m_{\ti t_2}^2, \mu_\eff^2) \; ,
\end{align}
with the effective $\mu$-parameter $\mu_{\eff} \equiv \lambda v_s/\sqrt{2}$, 
the Yukawa couplings $y_t\equiv \sqrt{2} m_t^{\DRb}(\mu_R)/v_u$
and $y_b\equiv \sqrt{2} m_b^{\DRb}(\mu_R)/v_d$ and  with the auxiliary function 
\be
I(a,b,c)= -\fr{ab\ln(a/b) + bc\ln(b/c)+ ca\ln(c/a)}{(a-b)(b-c)(c-a)}
\; .
\ee
\section{Loop-Corrected Higgs-to-Higgs Decays \label{sec:higgstohiggs}}
In this section we present the calculation of the loop-corrected
partial decay widths of all kinematically allowed Higgs boson decays
into two lighter Higgs bosons, $H_i\to H_jH_k$, $H_i\to A_lA_m$ and
$A_l\to A_m H_i$ $(i,j,k=1,2,3, \, l,m=1,2)$.  The two-body
decay width of a scalar $a$ decaying into two scalars $b$ and $c$ is given by
\beq
\Ga(a\to bc) =  R\, \fr{\la^{1/2}(m_a^2, m_b^2,m_c^2)}{16\pi m_a^3} \, 
|{\cal M}_{a\to bc}|^2, 
\eeq
where $R =1/2!$ for two identical final state particles and $R=1$
otherwise. The decay amplitude is denoted by ${\cal M}_{a\to bc}$ and 
\be
\la(x,y,z)= x^2+y^2+z^2-2xy-2xz-2yz \; .
\ee

In order to calculate the decay amplitude at one-loop order, one has to take
into account that not only the masses of the Higgs bosons receive
corrections, but also the fields themselves are affected. 
In the $\DRb$ scheme, which we are using for the Higgs field
renormalisation, the residue of the Higgs boson propagators is not
equal to one so that finite wave-function renormalisation factors $\bfZ$ have
to be taken into account in order to ensure the on-shell properties of
external Higgs bosons \cite{Dabelstein:1995js,Frank:2006yh}. The
transformation of the interaction states $h_u,h_d,h_s$ and $a,a_s$, respectively, to
the loop-corrected mass eigenstates, which we denote by capital
letters, $H_1,H_2,H_3$ and $A_1,A_2$, is then performed by radiatively
corrected transformation matrices for the scalar and pseudoscalar
sector, ${\cal R}^{S, l}$, ${\cal R}^{P,l}$, which are given by
\beq
{\cal R}^{S,l}_{is} &=& (\bfZ^S)_{ij} {\cal R}^S_{js} \; , \qquad
i,j,s=1,2,3 \;, \\
{\cal R}^{P,l}_{is} &=& (\bfZ^P)_{ij} {\cal R}^P_{js} \; , \qquad
i,j,s=1,2 \;. 
\eeq
They are built up by the finite scalar and pseudoscalar wave-function
renormalisation factors, $\bfZ^S$ and $\bfZ^P$, 
and by the respective rotation matrix ${\cal R}^S$, ${\cal R}^P$,
performing the rotation from the interaction states to the mass
eigenstates at tree-level, as defined in \eq{eq:rotationS} and
\eq{eq:rotationP}. Hence, for the scalar case, the indices correspond
in ascending order to the following Higgs entries: $i\mathrel{\widehat{=}} H_1,H_2,H_3$,
$j\mathrel{\widehat{=}} h_1,h_2,h_3$, $s\mathrel{\widehat{=}} h_d,h_u,h_s$.
And for the pseudoscalar case: $i\mathrel{\widehat{=}} A_1,A_2$,
$j\mathrel{\widehat{=}} a_1,a_2$, $s\mathrel{\widehat{=}} a,a_s$.
The wave-function renormalisation factor matrices are given by \cite{Williams:2007dc}
 \beq 
\bfZ^{S} = \begin{pmatrix}
             \sqrt{\hat{Z}_{H_1}} & \sqrt{\hat{Z}_{H_1}}
             \hat{Z}_{H_1H_2}& \sqrt{\hat{Z}_{H_1}} \hat{Z}_{H_1H_3}\\
              \sqrt{\hat{Z}_{H_2}} \hat{Z}_{H_2H_1}&
              \sqrt{\hat{Z}_{H_2}}& \sqrt{\hat{Z}_{H_2}} \hat{Z}_{H_2H_3}\\ 
               \sqrt{\hat{Z}_{H_3}} \hat{Z}_{H_3H_1}&
               \sqrt{\hat{Z}_{H_3}} \hat{Z}_{H_3H_2}&\sqrt{\hat{Z}_{H_3}}
            \end{pmatrix} \;,
\eeq
for the scalar Higgs bosons and by 
\beq
 \bfZ^{P} = \begin{pmatrix}
             \sqrt{\hat{Z}_{A_1}} & \sqrt{\hat{Z}_{A_1}} \hat{Z}_{A_1 A_2}\\
              \sqrt{\hat{Z}_{A_2}} \hat{Z}_{A_2 A_1}& \sqrt{\hat{Z}_{A_2}}
            \end{pmatrix} \;, \label{eq:neutralHiggs_WF} 
\eeq
for the pseudoscalar sector, with
\beq
\hat{Z}_i = \frac{1}{\left( \frac{i}{ \Delta_{ii} (p^2)}
  \right)^\prime (M_i^2)} \qquad \mbox{and} \qquad
\hat{Z}_{ij} = \left. \frac{ \Delta_{ij} (p^2)}{ \Delta_{ii} (p^2)}
\right|_{p^2=M_i^2} 
\label{eq:zfactor}
\eeq
for CP-even Higgs bosons $i,j=H_1,H_2,H_3$ and for CP-odd Higgs bosons
$i,j=A_1,A_2$. The diagonal $\Delta_{ii}$ and the off-diagonal
$\Delta_{ij}$ are given by the matrix elements of the 
two-point vertex function matrices for the scalars, $\hat{\Gamma}^S$,
Eq.~(\ref{eq:mass_Higgs_matrix}), and the pseudoscalars,
$\hat{\Gamma}^P$, Eq.~(\ref{eq:massmatrix}), as
\beq
\Delta^S = - \left[\hat{\Gamma}^S (p^2) \right]^{-1} \;, \quad
\Delta^P = - \left[\hat{\Gamma}^P (p^2) \right]^{-1} \;.
\eeq
The prime in Eq.~(\ref{eq:zfactor}) denotes the derivative with
respect to $p^2$. Note also that at one-loop order the complex eigenvalues of the
loop-corrected two-point vertex functions are used
in the evaluation of the wave function renormalisation factors, {\it
  i.e.}~$M_i$ in Eq.~(\ref{eq:zfactor}) includes also imaginary
parts. The mixing matrix elements in this approach therefore include the
full momentum dependence and imaginary parts of the one-loop Higgs
boson self-energies. The 
evaluation of the wave function renormalisation factors at zero
external momentum, $p^2=0$, which correspond to the result in the
effective potential approximation, on the other hand leads to a
unitary mixing matrix. \s

\begin{figure}[ht]
  \centering
  \includegraphics[width=0.9\textwidth]{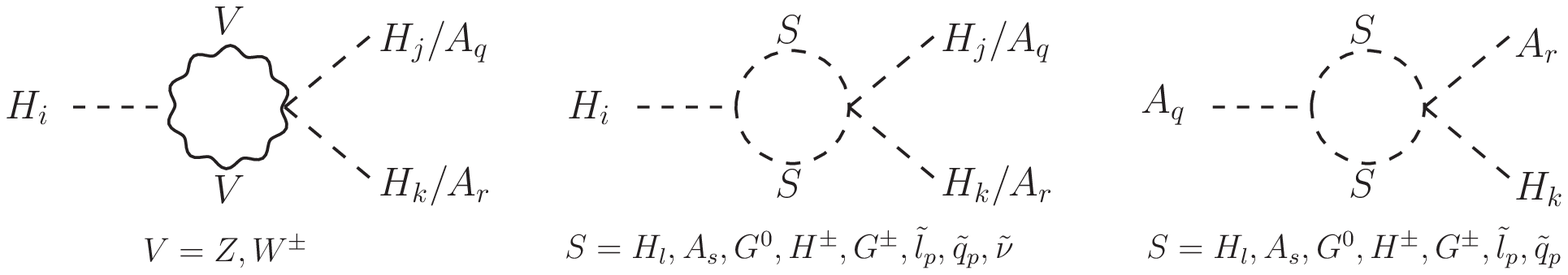}
  \\[0.5cm]
  \includegraphics[width=0.95\textwidth]{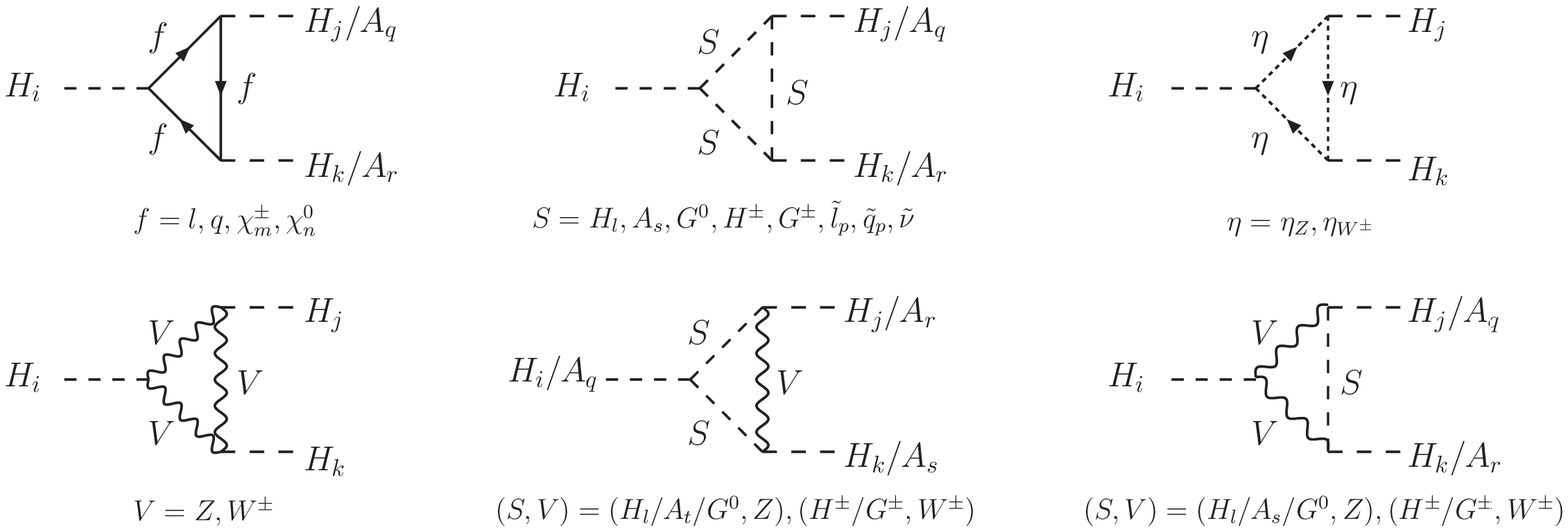}
  \caption{Generic Feynman diagrams contributing to the 1-point
    irreducible vertex functions. They are grouped by loops over scalars
    ($S$), vector bosons ($V$), fermions ($f$) and ghost particles ($\eta$).}
 \label{fig:1pidiags}
\end{figure}

With these definitions, the amplitudes of Higgs boson decays at higher
order can then be written as follows ($i,j,k=1,2,3$, $l,m=1,2$) 
\begin{align}
{\cal M}_{H_i\to H_j H_k}&=\sum_{i',j',k'=1}^{3}\bfZ_{ii'}^S\bfZ_{jj'}^S\bfZ_{kk'}^S(\la_{h_{i'}
  h_{j'} h_{k'}} +\de M^{\text{1PI}}_{h_{i'} h_{j'} h_{k'}}) \;, \label{eq:amp1}\\
{\cal M}_{H_i\to A_l A_m}&=\sum_{i'=1}^{3}\sum_{l',m'=1}^2\bfZ_{ii'}^S\bfZ_{ll'}^P\bfZ_{mm'}^P
(\la_{h_{i'} a_{l'} a_{m'}} +\de M^{\text{1PI}}_{h_{i'} a_{l'}
  a_{m'}}) +\de M^{G,Z \text{mix}}_{H_{i} \to A_{l} A_{m}} \;, \label{eq:amp2}\\
{\cal M}_{A_l\to A_m H_i}&=\sum_{i'=1}^{3}\sum_{l',m'=1}^2 
\bfZ_{ii'}^S\bfZ_{ll'}^P\bfZ_{mm'}^P (\la_{h_{i'} a_{l'} a_{m'}} +\de
M^{\text{1PI}}_{a_{l'} a_{m'} h_{i'}})+\de M^{G,Z \text{mix}}_{A_l\to
  A_m H_i} \;,\label{eq:amp3}
\end{align}
where $\la_{h_{i'} h_{j'} h_{k'}}$, $\la_{h_{i'} a_{l'} a_{m'}}$ are the trilinear Higgs
couplings at tree-level. Their explicit expressions are given in
\appen{app:1}.  The 1-point irreducible (1PI) contributions to the
vertex functions are denoted by $\de
M^{\text{1PI}}_{abc}$. Generic diagrams are shown in
Fig.~\ref{fig:1pidiags}. They are built up by two- and three-point
functions. The two-point functions involve four-point vertices between
two Higgs bosons and two scalars (Goldstone bosons, Higgs bosons,
sleptons, squarks, sneutrinos) as well as four point vertices between
two Higgs bosons and two gauge bosons ($Z$ or $W^\pm$), see first row
of Fig.~\ref{fig:1pidiags}. The three-point functions are given by loops
over scalars, gauge bosons, fermions as well as ghost particles
($\eta=\eta_Z, \eta_{W^\pm}$), see second and third row of 
Fig.~\ref{fig:1pidiags}. In addition to these diagrams the
counterterms to the tree-level Higgs couplings are included in $\de M^{\text{1PI}}$.\s

The $\de M^{G,Z \text{mix}}_{a\to bc}$ stands for the
sum of the contributions from the mixing of the CP-odd
Higgs boson with the Goldstone ($G$) boson and with the $Z$ boson, respectively.
\begin{figure}[hb]
  \centering
  \includegraphics[width=0.6\textwidth,height=0.18\textwidth]{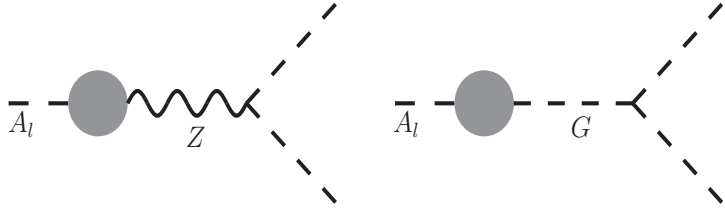}
  \caption{Generic one-loop Feynman diagrams involving $A_l Z$ and $A_l
    G$ transitions contributing to $\de M^{G,Z \, \text{mix}}$.}
 \label{fig:AG_AZtransitions}
\end{figure}
We use tree-level masses for the Higgs bosons in the loops in order to
ensure the proper cancellation of the UV-divergent pieces. But we use
the loop-corrected Higgs boson masses for 
the external particles in the evaluation of the wave-function
renormalisation factors, amplitudes and decay widths. While this does not affect
the UV-finite property of these quantities, 
the use of the loop-corrected Higgs boson masses breaks gauge
invariance in the decay processes involving CP-odd Higgs bosons, $H_i\to A_lA_m$ and
$A_2\to A_1 H_i$. The decay widths at one-loop level
contain contributions from Feynman diagrams involving $A_l Z$ and $A_l
G$ transitions as depicted in \fig{fig:AG_AZtransitions}. 
For these contributions the following Ward Slavnov-Taylor
identity ($l=1,2$) exists,
\beq
\hat\Si_{a_l G}(p^2) + \fr{ip^2}{M_Z} \hat\Si_{a_l Z}(p^2)= \left( p^2-
m_{a_l}^2 \right)\bigg({\cal R}^P_{l1} \, f_0(p^2)+ \fr12\sin2\beta\fr{\de
  \tan\beta}{\tan\beta} \, {\cal R}^P_{l1} - \fr12 \de
Z_{a_lG}\bigg)  \; , \label{eq:identity} 
\eeq
where $(i=1,2,3)$
\beq
f_0(p^2)&=& \fr{\al}{4\pi s_{2W}^2}\sum_{i=1}^{3} \left[ \cos2\beta \, {\cal
  R}^S_{i1} {\cal R}^S_{i2}  + 
 \cos\beta\sin\beta \, (({\cal R}^{S}_{i2})^2 - ({\cal R}^{S}_{i1})^2 ) 
\right] \, B_0(p^2, M_Z^2, m_{h_i}^2), 
\eeq
with $B_0(p^2, M_Z^2, m_{h_i}^2)$ denoting the scalar two-point function,
$\hat\Si_{a_l G}(p^2)$ and $\hat\Si_{a_l Z}(p^2)$ the renormalised
self-energies and $m_{a_l}$ and $m_{h_i}$ the CP-odd
and CP-even tree-level Higgs boson masses. The rotation matrices ${\cal R}^S$ and
${\cal R}^P$  have been defined in \eq{eq:rotationS} and
\eq{eq:rotationP}, respectively. By $\de \tan\beta$ we denote the
counterterm of $\tan\beta$ and by $\de Z_{a_lG}$ the wave function
counterterms. We have computed these identities using the same method
as in~Ref.~\cite{Baro:2008bg} and checked them numerically at arbitrary
momentum. We use them to test gauge invariance by applying
the general $R_\xi$ gauge for the
propagators of the exchanged $Z$ and Goldstone 
bosons. When $ p^2$ is set equal to the loop-corrected mass squared,
the right-hand side of \eq{eq:identity} does not vanish any more and
we therefore get a contribution to the amplitude from the mixing of
the pseudoscalar 
bosons with the $Z$ and the Goldstone boson, which depends on the
gauge fixing parameters. In order to get a gauge invariant amplitude,
one can use the tree-level masses for the CP-odd Higgs bosons
in the $A_l Z$ and $A_l G$ mixing diagrams, which has been applied in 
\cite{Williams:2007dc,Williams:2011bu}. Alternatively, one can use the
loop-corrected masses for the external particles also in these
contributions, which are then computed in 
the unitary gauge. We have applied both methods. The
difference between the two results is of higher order. In the end the
$A_l Z$ and $A_l G$ mixing contributions are small compared to the 
remaining contributions to the decay amplitude.  \s

For the determination of the loop-corrected Higgs boson masses, mixings and
trilinear Higgs boson self-couplings two independent calculations have been
performed. While in both calculations the
necessary model file was created using the program {\tt Sarah}
\cite{sarah}, one of them is based on a Fortran code that uses 
\favn~\cite{Kublbeck:1990xc} to generate the Feynman diagrams, the other
one uses {\tt FeynArts-3.5}. In both calculations the amplitudes are
evaluated with \fcvn~\cite{Hahn:1998yk}, and the numerical evaluation of
the loop-integrals is performed with the program {\tt LoopTools}
\cite{Hahn:1998yk}. The required counterterms of the Higgs boson sector
are supplied by two independent Mathematica programs, which determine
these in the course of the calculation of the loop-corrected masses of
the neutral Higgs bosons. 

\section{Numerical Analysis \label{sect-results}}
\subsection{Input Parameters and Constraints \label{sect-constraints}}
In our numerical analysis, we use the following SM parameters
\cite{Beringer:1900zz,Jegerlehner:2011mw} 
\begin{align}
\al(M_Z)&=1/128.962 \,, &\al_s(M_Z)&= 0.1184 \,, &M_Z&=91.1876\,
\gev\,, \crn
M_W&=80.385\,\gev \,,   &M_t&=173.5\,\gev \,,
&m^{\MSb}_b(m_b^{\MSb})&=4.18\,\gev \;. \label{eq:param1}
\end{align} 
The running strong coupling constant $\al_s$ is evaluated at two-loop
order in the calculation of the loop-corrected NMSSM Higgs boson
masses and of the Higgs pair production cross sections. 
The top quark pole mass ($M_t$) and the $\overline{\text{MS}}$
bottom quark mass
will be used to compute the running quark masses at the renormalisation
scale $\mu_R=M_{\text{SUSY}}$, as described at the end of \sect{sec:HiggsMassNMSSM}. 
The running bottom and top quark masses are then used in the evaluation
of the loop-corrected Higgs boson masses, mixings and decay widths.  The
light quark masses have only a small influence on the results. They
are chosen as 
\beq
m_u=2.5\,\mev\; , \quad m_d=5\,\mev\; , \quad m_s=95\, \mev \quad
\mbox{and} \quad m_c=1.27\, \gev \;. \label{eq:param2}
\eeq

\noindent 
Concerning the NMSSM sector, we set the soft SUSY breaking masses and trilinear
couplings of the third generation and the gaugino mass parameters as
follows 
\begin{align}
 M_{\ti L_3}&=M_{\ti \tau_R}=250\, \gev,
\quad A_t=A_b=A_\tau=1.5\,\tev,\quad M_{\ti Q_3}=M_{\ti t_R}=M_{\text{SUSY}},\crn
 M_{\ti b_R} &=1\,\tev,\quad M_1 = 162\,\gev,\quad M_2= 340\,\gev,
 \quad M_3=1\,\tev \; . \label{eq:param3}
\end{align}
The soft SUSY breaking masses and trilinear couplings of the first and
second generations also only slightly affect our results and have
been set to
\beq
M_{\ti L_{1,2}} =M_{\ti e_R} = M_{\tilde{\mu}_R} = 2\;\mbox{TeV}\, , \;
A_{u,c} = A_{d,s} = 2\; \mbox{TeV} \, , \; M_{\tilde{Q}_{1,2}} =
M_{\tilde{u}_R} = M_{\tilde{c}_R} = 2\; \mbox{TeV} \;. \label{eq:param4}
\eeq
These values guarantee a supersymmetric particle spectrum which is
in accordance with present LHC searches for SUSY particles
\cite{lhcsusy}. \s

Over the remaining NMSSM parameters we perform a scan with the
following restrictions: The SUSY mass scale $M_{\text{SUSY}}$ which
controls the soft SUSY breaking masses of the third generation is
chosen such that we can have light stop and sbottom masses which
are still in accordance with the LHC exclusion limits
\cite{thirdlhc}. We vary it as
\beq
650\; \mbox{GeV } \le M_{\text{SUSY}} \le 750 \; \mbox{GeV} \;.
\eeq
This leads to stop and sbottom masses of 
\beq
469 \; \mbox{GeV } &\le& M_{\tilde{t}_1} \le 607 \; \mbox{GeV} \; , \quad
823 \; \mbox{GeV } \le M_{\tilde{t}_2} \le 902 \; \mbox{GeV} \; , \\
655 \; \mbox{GeV } &\le& M_{\tilde{b}_1} \le 752 \; \mbox{GeV} \; , \quad
1000.2 \; \mbox{GeV } \le M_{\tilde{b}_2} \le 1000.3 \; \mbox{GeV}  \;.
\eeq
The value of $\tan\beta$ is chosen as
\beq
2 \le \tan\beta \le 10 \;.
\eeq
Low values of $\tan\beta$ allow to maximize
the tree-level mass of the lightest Higgs boson so that the Higgs mass
corrections which are governed by the stop sector can be kept small
enough to avoid large fine-tuning \cite{King:2012is,King:2012tr}. Also
the effective parameter $\mu_{\eff}$ is taken as small as possible for
fine-tuning reasons, and is varied in the range
\beq
100 \; \mbox{GeV } \le \mu_\eff \le 200 \; \mbox{GeV} \;.
\eeq
To keep $\lambda$ and $\kappa$ in the perturbative regime up to the GUT
scale we choose
\beq
0 \le \lambda, \; \kappa \le 0.7 \qquad \mbox{with} \quad
\sqrt{\la^2+\kappa^2}<0.7 \;.
\eeq
The charged Higgs boson mass (which replaced the original parameter
$A_\lambda$) is varied in a range respecting the experimental exclusion
limits \cite{charmass}, 
\beq
160 \; \mbox{GeV } \le M_{H^\pm} \le 1 \; \mbox{TeV} \;.
\eeq
Finally, the tree-level mass of the lightest pseudoscalar Higgs boson, 
$m_{a_1}$, is chosen in the interval
\beq
0 \; \mbox{GeV } \le m_{a_1} \le 1 \; \mbox{TeV} \;. 
\eeq
The variation of $m_{a_1}$ instead of $A_\kappa$ allows a better
control over the mass of the singlet-like CP-odd Higgs boson, which can
be $A_1$ or $A_2$ depending on the parameter set. Among the points
that have been generated in the above parameter space, we selected
only those which satisfy the following constraints 
arising from the LHC discovery of a SM-like Higgs boson
\cite{:2012gk,:2012gu} and from the exclusion limits reported
by LEP, Tevatron and LHC:
\ben
\item[1.)] One of the scalar Higgs bosons $H_1$ or $H_2$, denoted by $h$ in
  the following, is demanded to have a loop-corrected mass in the range
\beq
124 \; \mbox{GeV } \le M_h \le \; 127 \; \mbox{GeV} \;. \label{eq:massrestr}
\eeq
\item[2.)] We check our parameter points for compatibility with the
  experimental best fit values to the signal strengths
  \cite{bestfitvalues1,bestfitvalues2}.  For this we define the quantity,
\beq
R_{XX} (h) = \frac{\sigma (gg \to h)}{ \sigma (gg \to H_{\text{SM}}) }
\times \frac{\text{BR}(h \to XX)}{ \text{BR}(H_{\text{SM}}\to XX) }
\equiv R_\sigma (h) \times R_{XX}^{\text{BR}} (h) \;, \label{eq:RXX}
\eeq
which measures the rate of an NMSSM Higgs boson $h$ with mass near
125~GeV, 
produced in gluon fusion and decaying into the final state $X$, compared to the
corresponding value of the SM Higgs boson $H_{\text{SM}}$ with same mass
  as $h$. Here, 
$\text{BR} (H \to XX)$ denotes the branching ratio of the decay of
the Higgs boson $H$ ($H=H_{\text{SM}}, h$) into the final state $XX$
and $\sigma$ is the production cross section via gluon
fusion. Since 
for a SM-like Higgs boson the main production mechanism is given by
gluon fusion, it is sufficient to restrict ourselves to gluon fusion
in the production. The SUSY and NMSSM Higgs boson particle
spectrum has been calculated with our own Fortran code, which also
calculates the loop-corrected Higgs boson 
masses that are needed {\it e.g.} for the external particles in the calculation
of the Higgs-to-Higgs decays, {\it
  cf.}~section~\ref{sec:higgstohiggs}. The branching ratios and partial
widths are evaluated with 
a Fortran code which we have written ourselves by modifying the program {\tt
HDECAY} \cite{Djouadi:1997yw,Djouadi:2006bz} to the case of NMSSM
Higgs bosons. We thus include the most important
higher order QCD corrections in the decay widths.\footnote{We have not
included electroweak corrections, as they cannot easily be transferred
from the SM/MSSM to the NMSSM. QCD corrections on the other hand
do not involve Higgs couplings, so that they can readily be taken over
for the NMSSM.} We use the ratio of
the partial decay width of $h$ into gluons with respect to the one of
the SM Higgs boson in order to approximate $R_\sigma (h)$ in
Eq.~(\ref{eq:RXX}).  \s

In the NMSSM, there can be scenarios where
two Higgs bosons are close in mass so that the signal is not built up
by a single Higgs boson but by a superposition of the rates of
neighbouring Higgs bosons, which depends of course on the experimental
resolution in the respective final state. In order to compare with the
experimentally measured signal strengths $\mu_{XX}$ in the various
final states, we introduce the reduced cross sections $\mu_{XX}$ which
are built up by the superposition of the rates from an NMSSM $h$ boson near
125~GeV and other NMSSM Higgs bosons $\Phi=H_i,A_l$ ($i=1,2,3,\;
l=1,2$) close in mass. It is given by
\beq
\mu_{XX} (h) \equiv R_{\sigma} (h) \, R_{XX}^{BR} (h) \;\; + \hspace*{-0.4cm}
\sum_{\scriptsize \begin{array}{c} \Phi\ne h \\ |M_{\Phi}\!-\!M_h| \le
  \delta \end{array}} \hspace*{-0.4cm} R_{\sigma} (\Phi)
\, R_{XX}^{BR} (\Phi) \, F(M_h, M_\Phi, d_{XX}) \;. \label{eq:mudef}
\eeq
By $\delta$ we denote the mass resolution in the respective
final state $XX$. The superposition with the non-$h$ Higgs bosons is weighted with a
Gaussian weighting function $F(M_h, M_\Phi, d_{XX})$. The
parameter $d_{XX}$, which influences the width of the weighting
function, takes into account the  experimental resolution of the
different channels.\footnote{We follow here the approach 
  implemented in the program package {\tt NMSSMTools}
  \cite{nmssmtools,webnmssm}, which is based on NMSSM extensions
of the Fortran codes {\tt HDECAY} \cite{Djouadi:1997yw,Djouadi:2006bz} and {\tt
  SDECAY} \cite{Djouadi:2006bz,Muhlleitner:2003vg}.} \s

In order to comply with the recent Higgs search results of the best
fits to the signal strengths in the $\gamma\gamma$ and massive gauge
boson $WW$, $ZZ$ final states, we only keep parameter points which
lead to a Higgs mass spectrum with the following conditions:
\beq
&&\hspace*{-1.6cm} 
\mbox{\underline{Conditions on the parameter scan:} } \nonumber \\[0.2cm]
&&\hspace*{-1.5cm}\begin{array}{ll}
\mbox{At least one CP-even Higgs boson $h$ with: } & 124 \mbox{ GeV }
\lsim M_{h} \lsim 127 \mbox{ GeV } \\[0.2cm]
\mbox{For } 124 \mbox{ GeV } \!\lsim\! M_{h}= M_{H_{\scriptsize
    \mbox{SM}}} \!\lsim\! 127 \mbox{ GeV } \\[0.1cm]
\mbox{the reduced cross sections for $\gamma\gamma$ must fulfill: } &
\mu_{\gamma\gamma} (h) \ge 0.8 \\[0.1cm]
\mbox{the reduced cross sections for $ZZ$, $WW$ must fulfill: } 
& 0.8 \le \mu_{ZZ} (h), \,\mu_{WW} (h) \le 1.2 
\end{array} 
\label{eq:cond}
\eeq

For the rates in the $b\bar b$ and $\tau \bar \tau$ final states, we
do not apply any restriction since these channels suffer from large
uncertainties up to date.
\item[3.)] We use \higgsbound\, \cite{Bechtle:2008jh} to
  verify that the Higgs mass spectrum resulting from the respective chosen
  parameter set is allowed by the 
  published exclusion bounds from the Higgs searches at LEP, Tevatron
  and LHC.\footnote{We have included the latest results of the
    exclusion bounds in the dominant channels, {\it i.e.}
    $\gamma\gamma$ \cite{gamgam}, $ZZ$ \cite{zz} and $W^+W^-$
    \cite{ww}.} Otherwise the parameter set is rejected. 
\een

If not stated otherwise, in the following numerical analysis we keep only
those parameter sets of our parameter scan which fulfill the restrictions
1.)-3.). Furthermore, we call the NMSSM Higgs boson, which fulfills
the conditions Eq.~(\ref{eq:massrestr}) and Eq.~(\ref{eq:cond}), {\it i.e.}
which has a mass value around 125~GeV and rates compatible with the
LHC searches in the gauge boson final states, SM-like and denote it by
$h$. Note, however, that calling a Higgs boson SM-like according to
these definitions does not necessarily imply that it has SM-like
couplings. It is only the reduced signal strengths which we demand to be
SM-like.

\subsection{Effective trilinear Higgs couplings \label{sec:effective}}
Before we investigate the effect of the one-loop corrected trilinear
Higgs boson self-couplings on Higgs boson phenomenology, namely on
Higgs-to-Higgs decays and Higgs boson pair production, we discuss in
this subsection the effective trilinear Higgs couplings, that are
defined in the following. \s 

In the SM, the trilinear Higgs coupling at tree-level is given by 
\beq
\la_{SM}^\tree=\frac{3 M_{H}^2}{v} \;, 
\eeq
where $M_{H}$ is the physical SM Higgs boson mass and
$v=246$~GeV is the vacuum expectation value. In the NMSSM we
have
\beq
v=\sqrt{v_u^2+v_d^2} \;.
\eeq
We define the one-loop  corrected effective  trilinear SM Higgs coupling
as the one which includes the one-loop contributions evaluated at zero
external momenta. The calculation is performed by applying the on-shell
renormalisation scheme for $e, M_Z, M_W, M_{H}$ and the tadpole
 as well as for the Higgs field. For Higgs
boson masses below $160\,\gev$, the effective  trilinear Higgs
coupling can be approximated by 
\beq
\la_{SM}^\eff = \la_{SM}^\tree \braket{1 - \fr{\al M_t^4}{\pi M_{H}^2
    M_W^2 s_W^2}} \;,
\label{eq:hhhsmeff}
\eeq  
with $s_W \equiv \sin\theta_W$ and $\theta_W$ denoting the weak
angle. This coincides with the result of
Ref.~\cite{Kanemura:2004mg}. For a heavier Higgs boson, diagrams with
Higgs bosons inside the loops give important contributions of order ${\cal
  O}(M_H^4, M_H^2M_t^2)$. As we only consider SM Higgs boson masses
of 125~GeV, we can use the approximation Eq.~(\ref{eq:hhhsmeff}).
\s

The one-loop corrected effective  trilinear Higgs self-couplings of the
NMSSM are defined in the same way as the SM one. This means that the
expressions \eq{eq:amp1}-\eq{eq:amp3}, which are nothing else but the one-loop
corrected Higgs boson self-couplings, have to be evaluated at zero external
momenta, and the wave function renormalisation matrix, $\bfZ^{S/P}$, 
 has to be replaced by the rotation matrix $\bfZ^{S/P} (p^2=0)$ of the
 loop-corrected renormalised CP-even/odd Higgs boson mass matrices
 evaluated at zero external momentum, $\hat{M}_{S/P}^2(0)$. The latter
 have been defined in \eq{eq:mass_Higgs_matrix}/\eq{eq:massmatrix}. 
On the other hand, the tree-level  effective  trilinear Higgs
couplings of the NMSSM are the tree-level couplings dressed by the
loop-corrected rotation matrices at zero external momentum
$\bfZ^{S/P} (0)$. In the same way we have applied the loop-corrected
rotation matrices $\bfZ^{S/P} (0)$ in the couplings of the Higgs
bosons to the remaining SM particles and implemented these in our
modified version of the program {\tt HDECAY}, which we use to
calculate the branching ratios of the NMSSM Higgs bosons. With these
definitions the effective couplings are 
real. If not stated otherwise we will use them in the following in particular in the
computation of the Higgs pair  production processes. \s

\begin{figure}[h]
  \centering
\includegraphics[width=0.6\textwidth,height=0.60\textwidth]{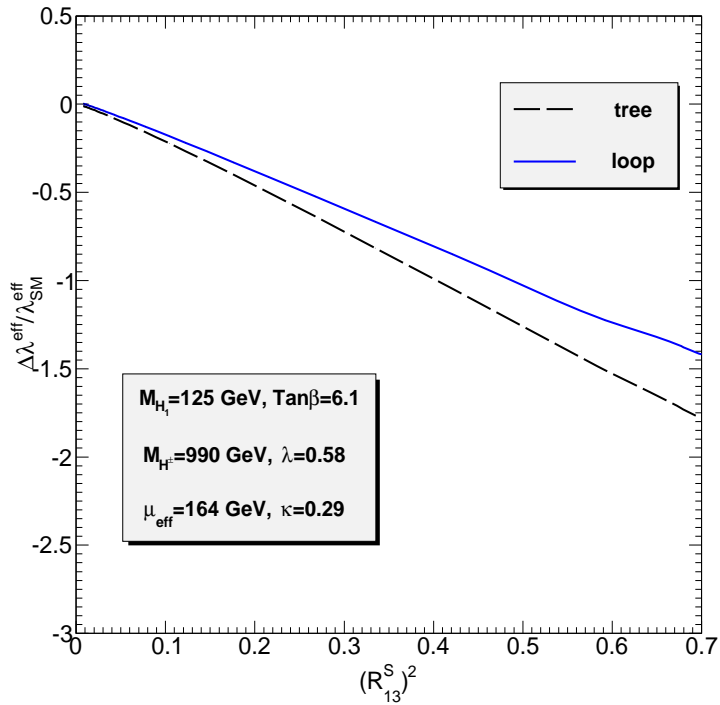}
  \caption{Normalised deviation of the trilinear Higgs self-coupling of the
    SM-like NMSSM Higgs boson, here $H_1=h$, from the corresponding SM
    coupling, $\Delta
    \lambda^\eff/\lambda^\eff_{\text{SM}}$, with $\Delta \lambda^\eff=
    \lambda^\eff_{hhh} -\lambda^\eff_{\text{SM}}$, as a
    function of the singlet admixture squared of $H_1$,
    $(R_{13}^{S})^2$, both for tree-level and one-loop corrected
    self-couplings. In this figure 
$A_t$ is varied such that $H_1$ is kept SM-like according to
our definitions in subsection~\ref{sect-constraints}. The remaining parameters
have been chosen as given in Eqs.~(\ref{eq:param1})-(\ref{eq:param4}). The $\DRb$
renormalised parameters are taken at the renormalisation scale,
$\mu_{R}=M_{\text{SUSY}}=700\,\gev$.} 
 \label{fig:HHH_coup}
\end{figure}

We first discuss the SM limit. In the framework of the MSSM, it has been
shown in 
\cite{Hollik:2001px,Dobado:2002jz} that both the tree-level and the
loop-corrected Higgs self-couplings (the trilinear and the quartic
one) of the lightest MSSM Higgs boson converge to  the corresponding SM ones
in the decoupling limit, \ie $M_{H^\pm} \gg M_Z$. This feature is not
shared by the NMSSM, also not at tree level, due to the mixing with
the singlet component. Figure~\ref{fig:HHH_coup} shows the difference
between the effective trilinear Higgs self-coupling of the lightest
NMSSM Higgs boson $H_1$ and the one of the SM Higgs boson. The 
difference is evaluated for the tree-level couplings normalised to the
tree-level SM effective coupling as well as for the 
one-loop couplings normalised to the one-loop SM effective coupling.
Hence ($H_1 \equiv h$)
\beq
\frac{\Delta \lambda^{\scriptsize \mbox{eff}}}{\lambda^{\scriptsize
    \mbox{eff}}_{\scriptsize \mbox{SM}}} = \left\{ \begin{array}{ll} 
\frac{\displaystyle \lambda^{\scriptsize
    \mbox{eff,tree}}_{hhh} - \lambda^{\scriptsize
    \mbox{eff,tree}}_{\scriptsize \mbox{SM}}}{\displaystyle \lambda^{\scriptsize
    \mbox{eff,tree}}_{\scriptsize \mbox{SM}}} & \mbox{ at tree-level}
\\[0.6cm]
\frac{\displaystyle \lambda^{\scriptsize
    \mbox{eff,1l}}_{hhh} - \lambda^{\scriptsize
    \mbox{eff,1l}}_{\scriptsize \mbox{SM}}}{\displaystyle \lambda^{\scriptsize
    \mbox{eff,1l}}_{\scriptsize \mbox{SM}}}  & \mbox{ at one-loop
  level} \;. 
\end{array} \right.
\eeq 
The normalised deviation is shown as a function of the singlet
admixture, which is given by the rotation matrix element ${\cal
  R}^S_{13}$. We have chosen a scenario where the lightest NMSSM Higgs
boson $H_1$ is SM-like according to our definition at the end of
\ssect{sect-constraints}. This means that it has a mass of about 125 GeV
and that the signal strengths in the gauge boson final states are
SM-like. Note, however, that this does not imply that $H_1$ has SM-like
couplings, in particular not for large singlet admixtures. The criteria
on the reduced signal strengths are in this case fulfilled only due to the
superposition with a second CP-even Higgs boson which is close in mass
and which has SM-like couplings. In
order to get different values for the mixing we have varied the
trilinear soft SUSY breaking coupling $A_t$ of the stop sector, while
taking care to keep $H_1$ at 125 GeV and to fulfill the constraints on
the reduced signal strengths. As can be inferred from the figure, $|\Delta
\lambda^\eff/\lambda^\eff_{\text{SM}}|$ increases with 
the mixing, both at tree-level and at one-loop level. It can become as
large as 1.4 at loop level, and as large as 1.75 at tree-level, provided that
$H_1$ is rather singlet-like.\footnote{In fact, with
  rising singlet admixture the CP-even Higgs 
  bosons $H_1$ and $H_2$, which are close in mass, interchange their
  roles. For large singlet admixtures, the deviation from the SM value
  of the trilinear Higgs 
  self-coupling at large masses $M_{H^\pm}$ is hence only an artefact of
  $H_1$ losing its role as Higgs boson with SM-like couplings.} 
In the limit of zero singlet admixture, $H_1 \equiv h$ becomes effectively
MSSM-like, and with the chosen large charged Higgs boson mass of
$M_{H^\pm}=990$~GeV we are in the SM-limit. In this SM-limit, the NMSSM
trilinear Higgs self-coupling coincides with the SM effective self-coupling,
as expected.
 
The one-loop corrected NMSSM effective  trilinear Higgs couplings
receive significant contributions from the diagrams with
(s)top quarks in the loops.\footnote{For small values of $\tan\beta$.}
There can be, however, extremely large contributions from the triangle
diagrams with 
light singlet-like bosons in the loop. In case $H_2$ is the
Higgs boson with mass around 125~GeV, $A_1$ and $H_1$ have to be
singlet-like in order to avoid the exclusion limits on light Higgs
boson masses. The analytic expressions for the one-loop contribution
$\delta \lambda^{A_1}$ and $\delta
\lambda^{H_1}$ of $A_1$, respectively $H_1$, running in the loop, 
can be cast into the form ($i=1,2,3$) 
\beq
\delta \lambda^{A_1}_{H_i H_i H_i} &=& \frac{1}{16 \, \pi^2} 
\left(\sum\limits_{j=1}^{3} {\bf Z}^{S}_{ij} \, \lambda_{h_{j} a_{1}
    a_{1}} \right)^3 \frac{1}{2 m_{a_1}^2} \;, \\ 
\delta \lambda^{H_1}_{H_i H_i H_i} &=& \frac{1}{16 \, \pi^2} \left(
  \sum\limits_{j=1}^{3} {\bf Z}^{S}_{ij} \, \lambda_{h_{j} h_{1}
    h_{1}} \right)^3 \frac{1}{2 m_{h_1}^2} \;,
\eeq
where we use tree-level masses $m_{a_1}$, $m_{h_1}$, respectively, for
the Higgs bosons inside the loops, and 
$\lambda$ denotes the tree-level trilinear couplings. 
These contributions hence blow up
\begin{figure}[hb]
  \centering
\includegraphics[width=0.45\textwidth]{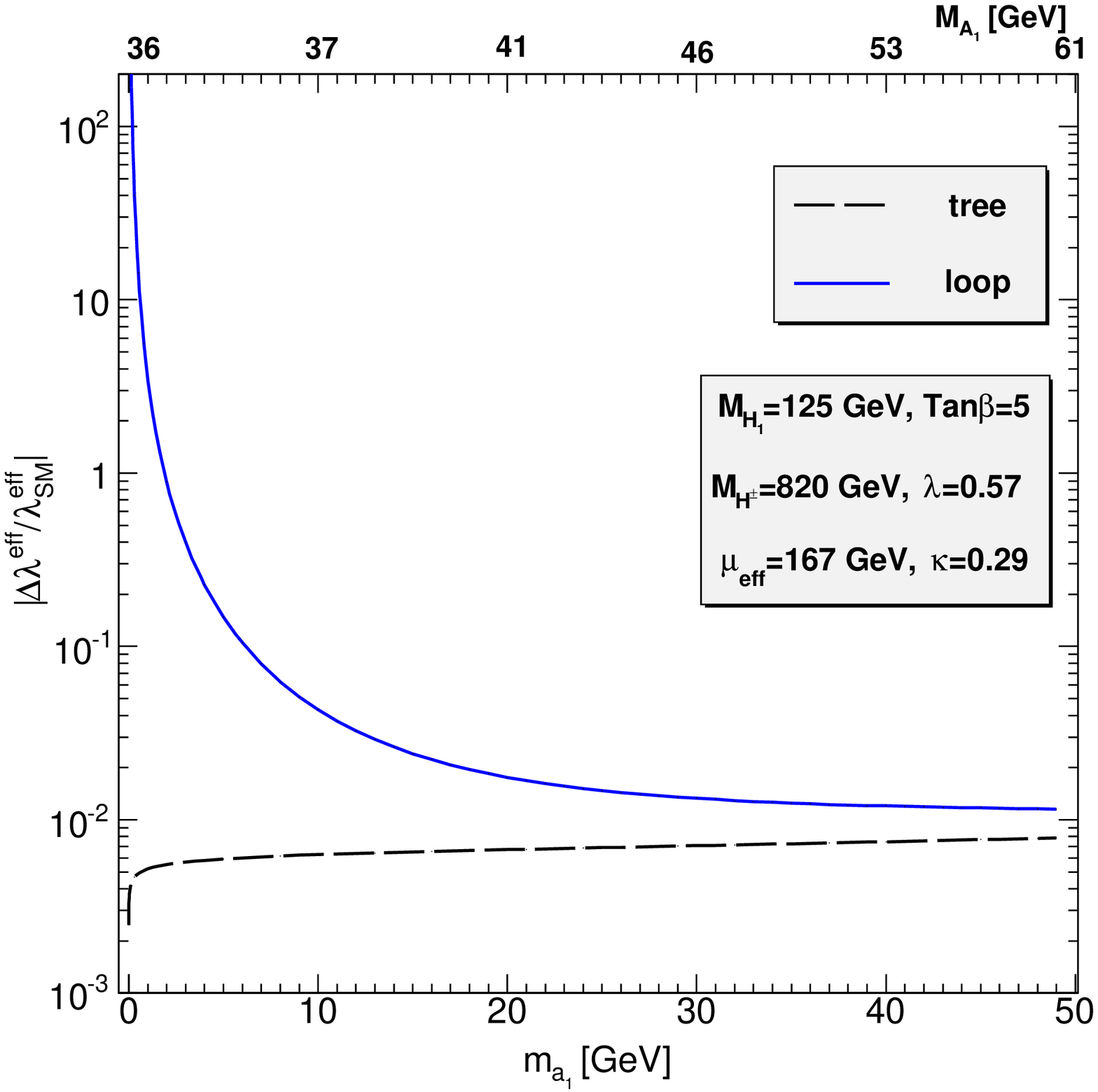}
\label{fig:A1re}\quad
\includegraphics[width=0.45\textwidth]{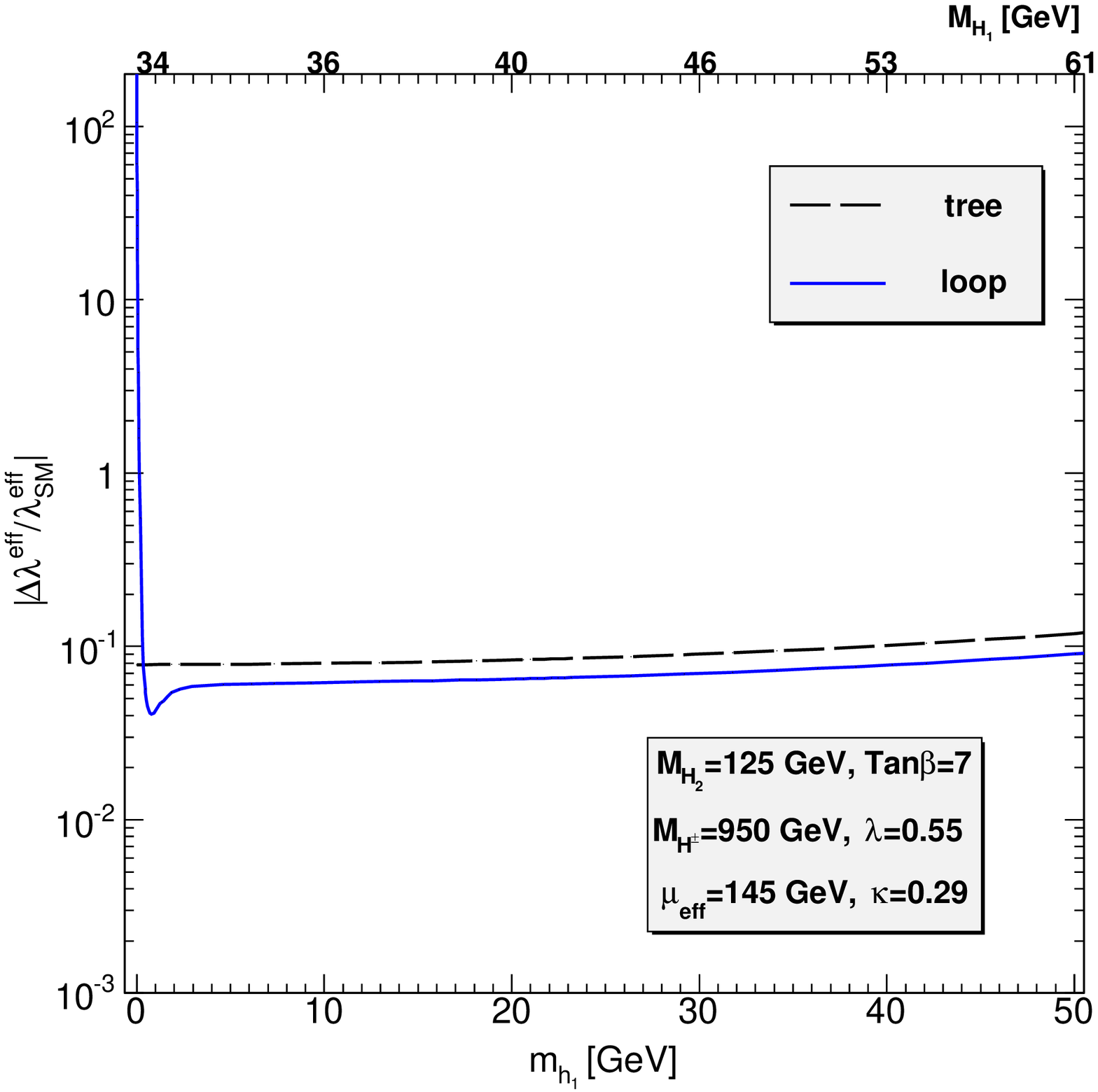}\label{fig:H1re}
  \caption{Absolute value of $\Delta
    \lambda^\eff/\lambda^\eff_{\text{SM}}$ for tree-level couplings
    (dashed) and one-loop corrected couplings (full) as a function of the
    tree-level lightest CP-odd Higgs mass $m_{a_1}$ for a
    125~GeV $H_1$ (left) and as a function of the
    tree-level lightest CP-even Higgs 
    mass $m_{h_1}$ for a 125~GeV $H_2$ (right). The scales on
    top of each figure refer to the loop-corrected mass $M_{A_1}$ (left) and
    $M_{H_1}$ (right). The parameters have been chosen as specified in
    Eqs.~(\ref{eq:param1})-(\ref{eq:param4}), and the renormalisation
    scale has been set $\mu_{R}=M_{\text{SUSY}}=700\,\gev$.} 
 \label{fig:HHH_A1H1}
\end{figure}
when $m_{a_1}$ or $m_{h_1}$ become small while the trilinear couplings
remain non vanishing.  We show examples of such large contributions to
$\la^{\eff}_{\text{NMSSM}}$ for $H_1$ having a mass around 125~GeV in
Fig.~\ref{fig:HHH_A1H1}~(left) and for a 125~GeV $H_2$ in
Fig.~\ref{fig:HHH_A1H1}~(right). In these figures we  use again the  
quantity $\Delta \lambda^\eff/\lambda^\eff_{\text{SM}}$ to
characterise the differences between the effective tree-level and one-loop
corrected NMSSM couplings and the corresponding SM ones. 
In Fig.~\ref{fig:HHH_A1H1}~(left) large contributions only arise from a light singlet-like 
$A_1$, as it is the only Higgs boson with mass smaller than
$H_1$. Figure~\ref{fig:HHH_A1H1}~(right) shows the case of   
$H_2$ having a mass near 125~GeV, so that large contributions 
can arise both from a very light $A_1$ and $H_1$. 
As can be inferred from the
figures, while the tree-level couplings are small, the loop
corrections become extremely large for light singlet masses. (The drop
of the tree-level coupling in Fig.~\ref{fig:HHH_A1H1}~(left) is due to
a cancellation between the various terms entering the coupling. The
kink in Fig.~\ref{fig:HHH_A1H1} (right) is due to a sign change in
$\Delta \lambda^{\scriptsize \mbox{eff}}$.) 
We note, however, that the existence of these huge
contributions  is not compatible with the constraints in
subsection~\ref{sect-constraints}, Eq.~(\ref{eq:cond}), on the 125~GeV
Higgs boson. The reason is the following. When the
tree-level mass $m_{a_1}$ ($m_{h_1}$) is very small, its
loop-corrected mass is  still small enough for the decay $H_1\to
A_1 A_1$ ($H_2\to H_1 H_1$) being kinematically possible. (The upper
scales of Figs.~\ref{fig:HHH_A1H1} show, respectively, the
loop-corrected masses $M_{A_1}$ and $M_{H_1}$.) These decays turn out to 
be important so that the branching ratios of the decays
into $\ga\ga, \, WW$ and $ZZ$ become very much suppressed and the decay
rates in these final states are not compatible any more with the best fit
values to these final state signal strengths given by the
experiment. A discussion on non-decoupling
effects of the one-loop trilinear couplings can also be found
in \cite{Kanemura:2010pa}, without, however, taking into account the
possibility of large singlet contributions.  

\subsection{Results for Higgs boson decays \label{sec:resdecays}}
In this subsection we show the effect of the one-loop corrected
trilinear Higgs self-couplings on the branching ratios of heavy Higgs
bosons into a pair of lighter Higgs bosons. These decays play a role in the 
\begin{figure}[ht]
  \centering
  \includegraphics[width=0.6\textwidth]{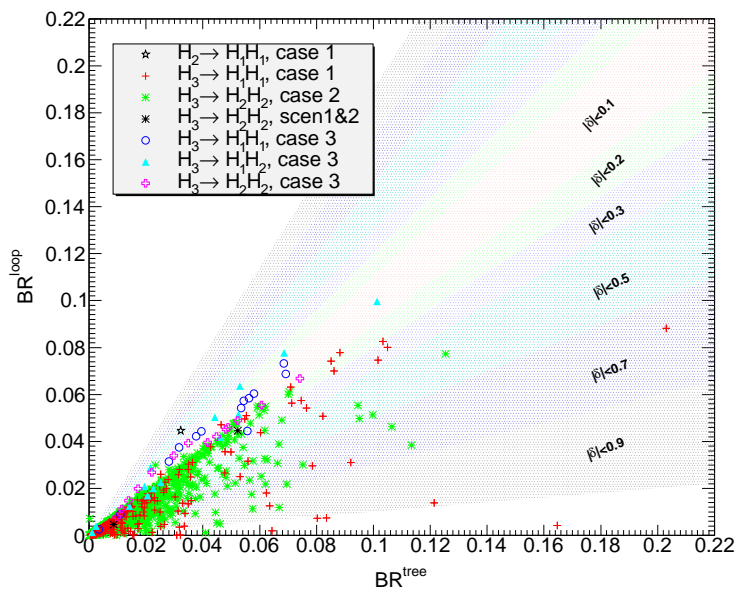}
  \caption{The branching ratio of heavier CP-even Higgs bosons, $H_2$
    or $H_3$, decaying into two SM-like Higgs bosons at loop
    level versus the tree-level branching ratio, for scenarios with
    $h\equiv H_1$ (case 1), $h\equiv H_2$ (case 2, scen1, scen2) as
    well as $H_1$ and $H_2$ close in mass near 125 GeV (case 3). 
    The difference between the one-loop and the 
    tree-level branching ratio is quantified by $\de \equiv
(\text{BR}^{\text{loop}}-\text{BR}^{\text{tree}})/\text{BR}^{\text{tree}}$. The
colored areas refer to different ranges of
$\de$.}
 \label{fig:HTO2HSM}
\end{figure}
search for the heavy Higgs bosons, which can possibly be detected at
the LHC via their decay into a pair of lighter Higgs bosons, which
then subsequently decay further into SM
particles. Figure~\ref{fig:HTO2HSM} shows the branching ratios of the 
decay of a heavy CP-even Higgs boson, $H_2$ or $H_3$ depending on the
scenario, into two SM-like Higgs bosons including loop-corrected
trilinear Higgs self-couplings, as defined in Eq.~(\ref{eq:amp1}),
versus the branching ratio evaluated with tree-level trilinear
couplings ({\it cf.}~Eq.~(\ref{eq:amp1}) without the 1-point
irreducible contributions). The three different cases refer
to scenarios with the SM-like Higgs boson being $H_1$, $h\equiv H_1$
(case 1), with $H_2$ being SM-like, $h\equiv H_2$ (case 2), and with
two light Higgs bosons $H_1$ and $H_2$ both having mass around 125~GeV
and the combined Higgs rates being compatible with the SM rates (case 3). As
can be inferred from the figure, which only includes scenarios
compatible with our constraints, the loop-corrected trilinear Higgs
coupling has the 
tendency to decrease the branching ratio compared to the tree-level
result. The deviations can be as large as 90\% in terms of the
tree-level branching ratio, in some cases even higher. Also shown by
black filled stars are two scenarios, where the SM-like Higgs boson is
given by $H_2$. The corresponding tree-level and loop corrected
branching ratios amount to
\beq
\begin{array}{ll}
(\mbox{BR}^{\scriptsize \mbox{loop}},
\mbox{BR}^{\scriptsize \mbox{tree}})_{H_3 \to H_2 H_2} = (4.7\times
10^{-3},8.7\times 10^{-3}) & \quad \mbox{scen1} \\
(\mbox{BR}^{\scriptsize \mbox{loop}},
\mbox{BR}^{\scriptsize \mbox{tree}})_{H_3 \to H_2 H_2} = (4.5\times
10^{-2},5.2\times 10^{-2}) & \quad \mbox{scen2} 
\end{array} \;.
\eeq
These two scenarios scen1 and scen2 have the characteristic feature of
being excluded if loop corrections are not taken into account. In
scen1 the tree-level branching ratio of the SM-like $H_2$ decay into
$H_1 H_1$ is as large as 0.02. This decreases the branching ratios into the
other SM particles so that the scenario would be excluded because the
reduced signal strength in the $WW$ final state drops below 0.8 and
the constraints Eq.~(\ref{eq:cond}) are not 
fulfilled any more. Only the inclusion of the loop corrections decreases
the branching ratio into $H_1 H_1$ to $2\cdot 10^{-4}$ so that the
scenario meets the constraints Eq.~(\ref{eq:cond}). In scen2 it is the branching ratio of
the decay $H_2 \to A_1 A_1$ which is as large as 0.26 at tree-level
compared to 0.1 after the inclusion of loop corrections, so that the
scenario is not valid at tree-level due to too small reduced signal strengths
in the $WW$ and $ZZ$ final states, unless loop 
corrections are included. These examples demonstrate the importance of
including the 
higher order corrections in order to properly interpret the NMSSM
spectrum with respect to the experimental findings. \s
  
\begin{figure}[h]
  \centering
\includegraphics[width=0.45\textwidth]{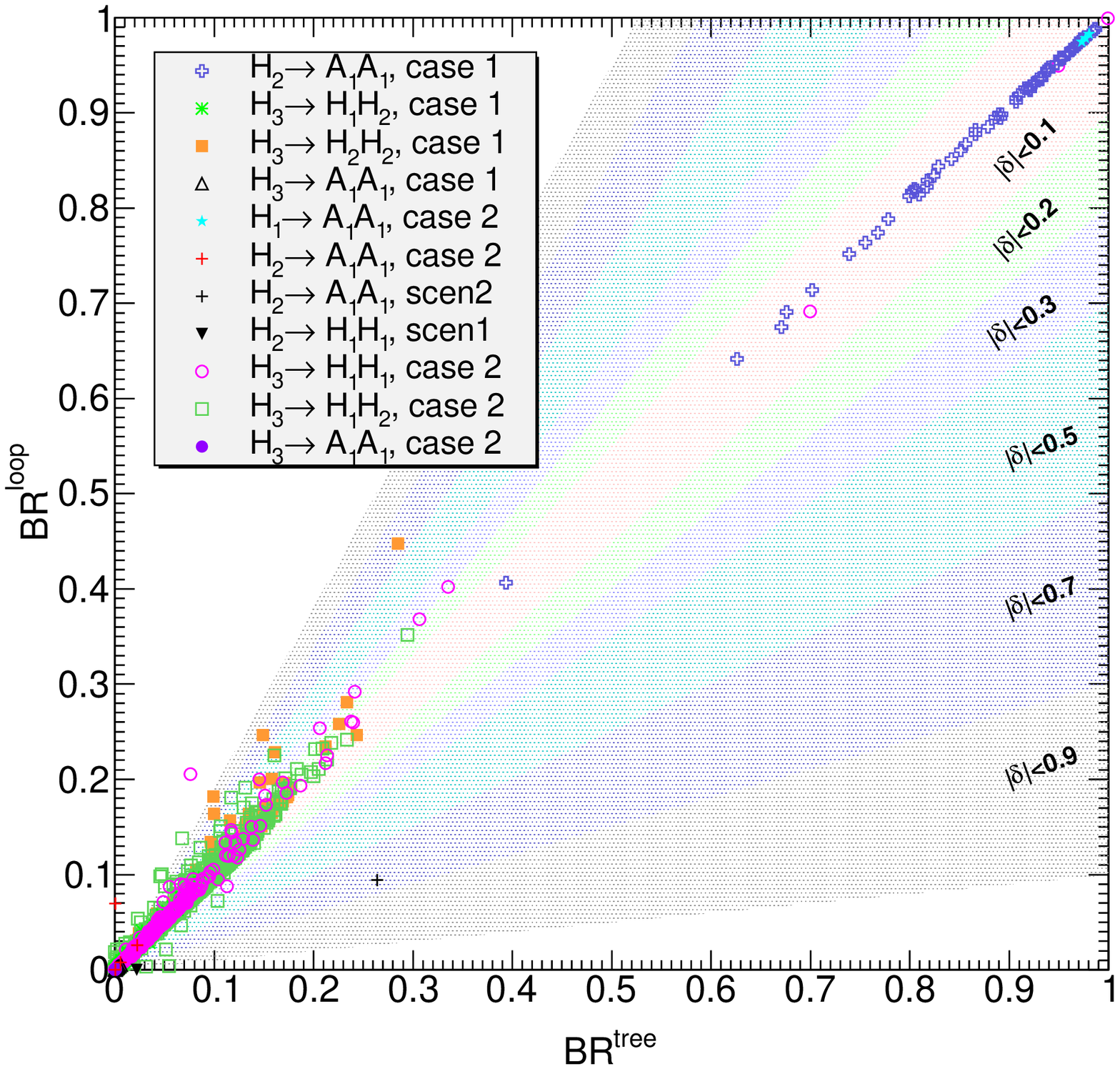}\label{fig:hdecay2}\quad
\includegraphics[width=0.45\textwidth]{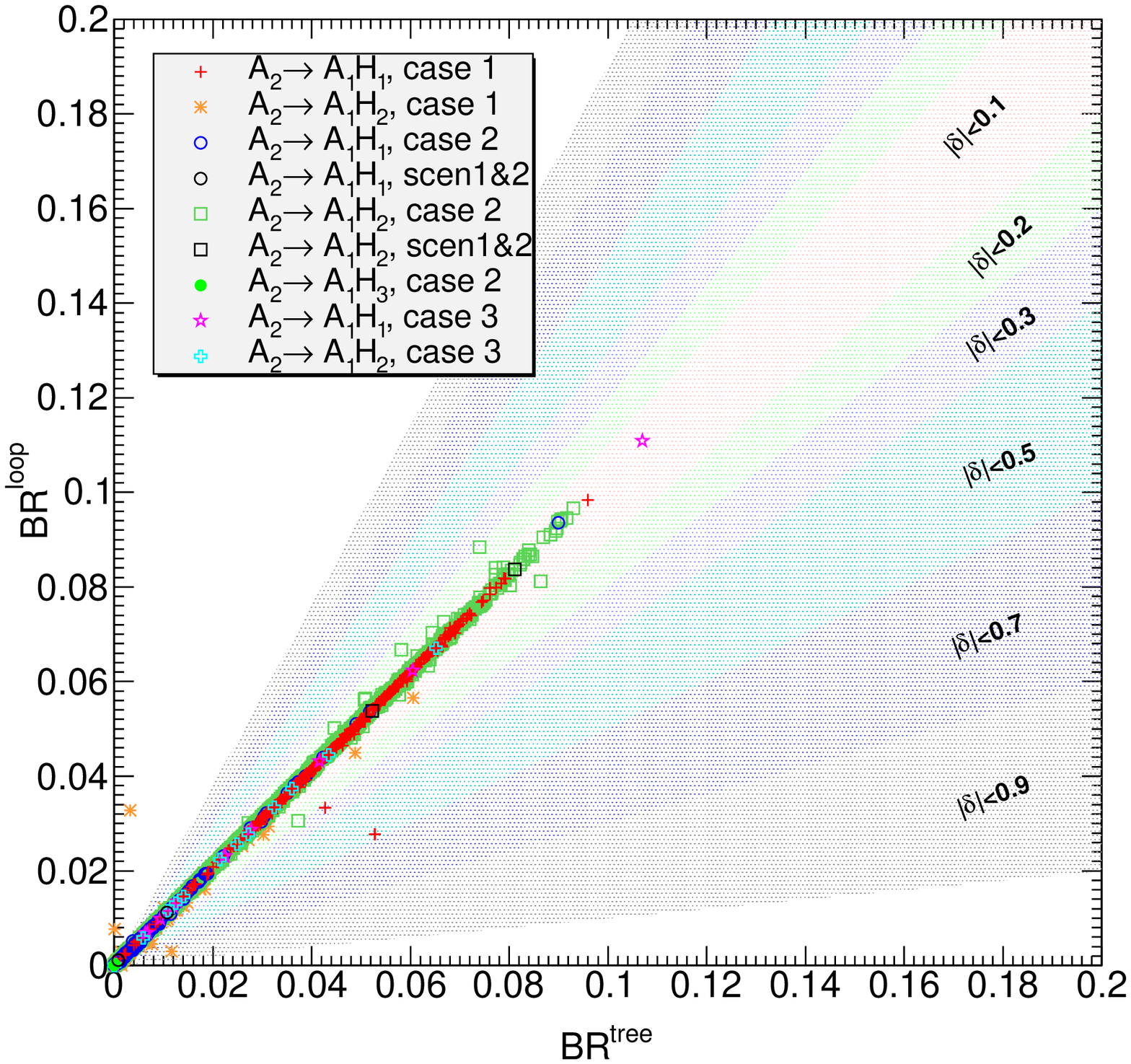}\label{fig:hdecay3}
  \caption{Same as \fig{fig:HTO2HSM} but for all other possible
    two-body decay modes of heavy scalar (left) and pseudoscalar
    (right) Higgs bosons into two lighter ones.}
 \label{fig:hdecay}
\end{figure}
In Fig.~\ref{fig:hdecay} left (right) we show the remaining possible branching
ratios of heavy scalar (pseudoscalar) Higgs bosons into a pair of
lighter Higgs bosons. In the left figure, the full black triangle corresponds to the
decay $H_2 \to H_1 H_1$ of scen1 and the black cross to the decay of
$H_2 \to A_1 A_1$ of scen2. The significant loop corrections leading to
deviations of up to 90\%, respectively 70\%, in terms of the
corresponding tree-level decay are crucial for the two scenarios to be
still in accordance with the LHC constraints Eq.~(\ref{eq:cond}). In
the right figure, the black squares are the branching ratio values of
the decay $A_2 \to A_1 H_2$ in scen1 and scen2, respectively. In this
decay the deviations between the tree-level and loop-corrected branching
ratios are small, less than 0.1 for both scenarios. \s

\begin{figure}[h]
  \centering
 \includegraphics[width=0.46\textwidth]{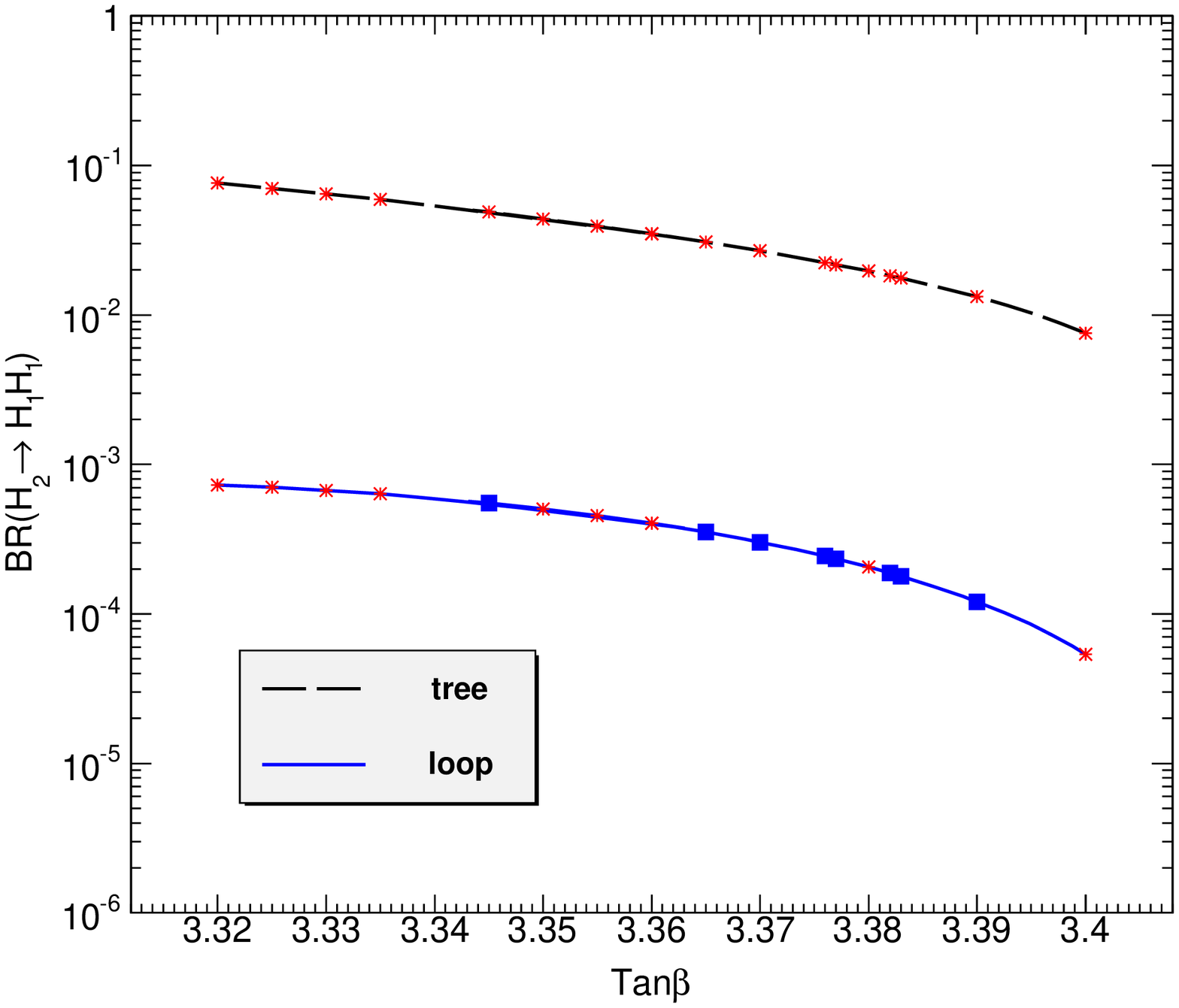} \hspace*{0.2cm}
  \includegraphics[width=0.46\textwidth]{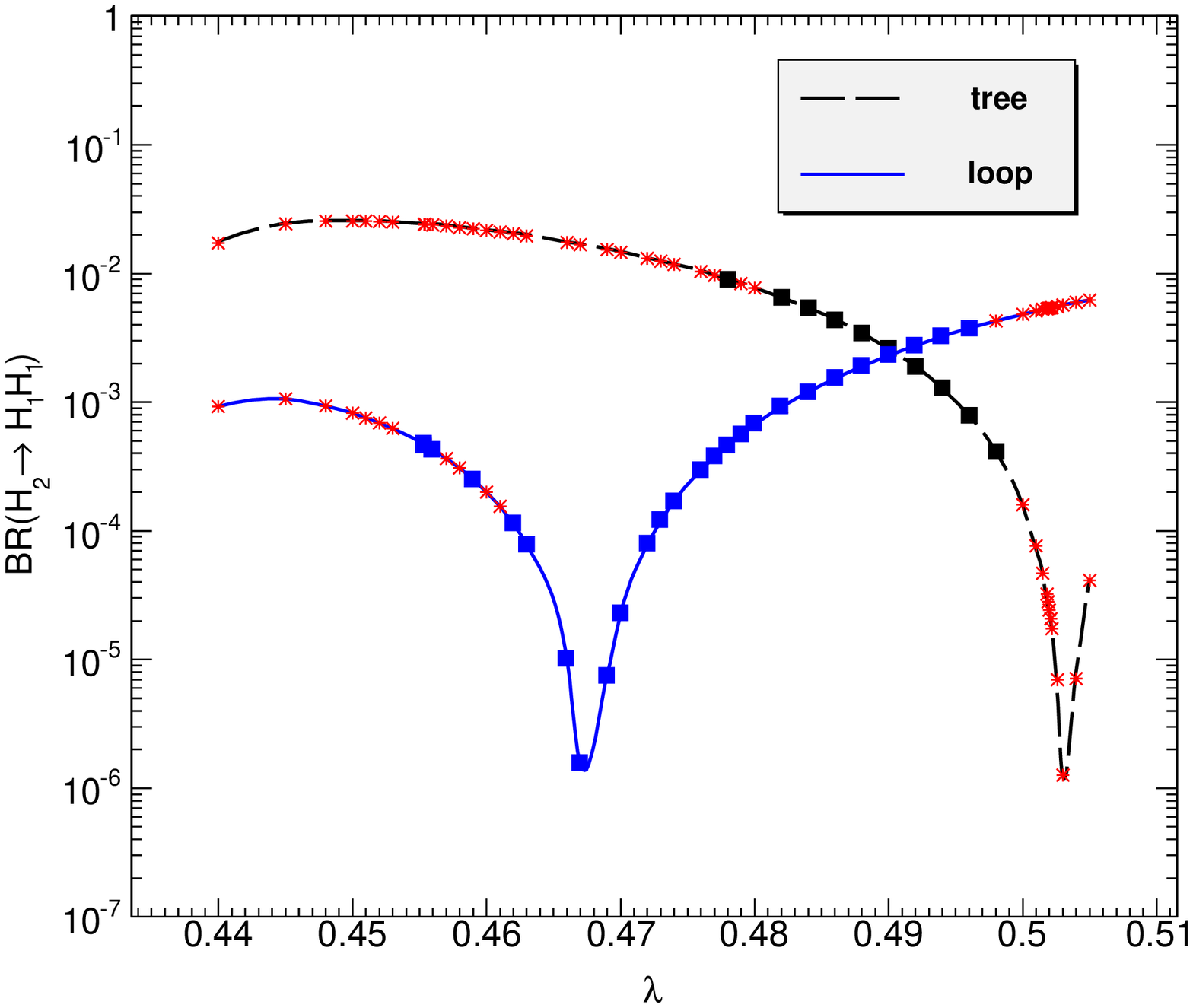}
  \caption{Branching ratio of $H_2$ with mass around 125~GeV into two
    light Higgs bosons $H_1$, at tree-level (dashed/black) and including
    loop corrections (full/blue) as a function of $\tan\beta$ (left) and of
  $\lambda$ (right). Lines are there to guide the eye. Box (star) points:
  allowed (excluded) due to the constraints Eq.~(\ref{eq:cond}).}
 \label{fig:H2H1H1}
\end{figure}

Figure~\ref{fig:H2H1H1} shows the branching ratio of the SM-like
$H_2$ decay into $H_1H_1$ as a function of $\tan \beta$ (left) and of
$\lambda$ (right), evaluated at tree-level and at loop level. All the
box points are allowed points, while the red star points do not
fulfill our constraints. The figure shows that there are scenarios
that would be excluded at tree-level, but are allowed once loop
corrections are taken into account, which suppress the non-SM decays into 
Higgs boson pairs, so that the branching ratios into the SM particles
remain compatible with the experimental constraints. The point of
scen1 in Fig.~\ref{fig:HTO2HSM} corresponds to $\tan\beta = 3.37$ in
Fig.~\ref{fig:H2H1H1}~(left) and to $\lambda=0.46$ in Fig.~\ref{fig:H2H1H1}~(right).
The plots illustrate once more the importance of loop corrections when
considering a specific model in light of the experimental results. As can be inferred
from the figure, the loop corrections not only change the absolute
value of the branching ratio but also shift the minimum to different
$\lambda$ values. The corresponding plots for a variation of the other
parameters $\kappa, A_\kappa, \mu_{\scriptsize \mbox{eff}}$ or
$M_{H^\pm}$ show a similar behaviour. 

\subsection{Pair production of neutral Higgs bosons at the
  LHC \label{sec:pairprod}} 

We apply the results of our calculation of the loop-corrected
Higgs boson self-couplings to the pair production of neutral Higgs bosons at the
LHC in order to study the effects of the higher order
corrections. Higgs pair production processes are important as they
give access to the trilinear Higgs self-couplings. The measurement of
the trilinear and quartic Higgs self-interactions allows for the
reconstruction of the Higgs 
potential which represents the ultimate check in the program of the
experimental verification of the Higgs mechanism
\cite{Djouadi:1999gv,Djouadi:1999rca}. The main
contribution to Higgs pair production at the LHC comes from $gg$
fusion \cite{Glover:1987nx,Plehn:1996wb,Dawson:1998py}.
Higgs pair production processes through vector boson fusion\cite{vbfhh}, double
Higgs-strahlung \cite{hradhh} and associated production of a Higgs
boson pair with $t\bar{t}$ \cite{tthh} are less important \cite{Baglio:2012np} so that we
focus here on the gluon fusion channel, namely the production of a pair of
SM-like Higgs bosons in the framework of the NMSSM.

\subsection{SM-like Higgs boson pair production through
  \boldmath{$gg$} fusion}
\begin{figure}[h]
  \centering
\includegraphics[width=0.8\textwidth]{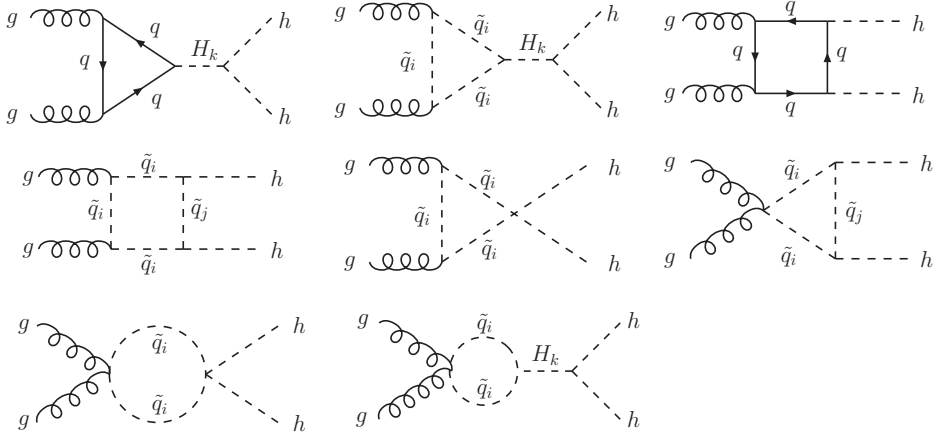}
  \caption{Generic diagrams contributing to pair production of a SM-like
    NMSSM Higgs boson $h$ in gluon fusion. The loops involve top and
    bottom (s)quarks, $q=t,b$, $\tilde{q}=\tilde{t},\tilde{b}$, $i,j=1,2$. The
    $s$-channel diagrams proceed via $H_k=H_1,H_2,H_3$, with one of
    these being the SM-like $h$, depending on the parameter choice.}
 \label{fig:doubleh}
\end{figure}

At leading order, the gluon fusion process into two SM-like Higgs
bosons $H_i \equiv h$ ($i=1,2$ depending on the chosen parameter set), 
\beq
gg \to H_i H_i \; , \qquad H_i \equiv h \;,
\eeq 
consists of triangle, box and two-point contributions mediated by top
and bottom (s)quarks, {\it cf.}~Fig.~\ref{fig:doubleh}.
The diagrams are similar to the MSSM case
\cite{Plehn:1996wb,Dawson:1998py}, except for the ones involving a
scalar Higgs boson in the $s$-channel, which gets contributions from all
possible three neutral NMSSM Higgs bosons, that subsequently split
via the corresponding trilinear Higgs coupling into the final state
Higgs pair. Thus the $hh$ production involves the trilinear Higgs 
self-couplings $\lambda_{hhh}$ and $\lambda_{hh H_k\ne h}$ ($k=1,2,3
\ne i$). The NLO QCD
  corrections have been calculated both for the SM and the MSSM case
  in the limit of heavy loop particle masses. The corrections are
  large, of ${\cal O}(100\%)$ \cite{Dawson:1998py}, and therefore have
  to be taken into account for a realistic assessment of the process. 
We use the public Fortran code {\tt HPAIR} \cite{HPAIR} which computes
the QCD corrected 
gluon fusion processes into pairs of Higgs bosons within the SM and
the MSSM, by applying the low-energy theorem. As claimed in \cite{Dawson:1998py}, 
this approximation is good for small values of $\tan\beta$ in the MSSM
and for Higgs boson masses below $2M_t$.\footnote{For further
  discussion of the application of the low-energy theorem to single
  and double Higgs production, see \cite{Gillioz:2012se}.} This is the
case for the here 
considered pair production of a SM-like NMSSM Higgs boson, hence with 
mass of about 125~GeV. Furthermore, in the NMSSM small $\tan\beta$
values are preferred in this case.  We have modified the code {\tt
  HPAIR} to include the additional contributions to the triangle
diagrams from the enlarged Higgs sector of the NMSSM and by replacing
the involved Higgs Yukawa and trilinear Higgs couplings with the
appropriate NMSSM Higgs couplings as discussed in subsection \ref{sec:effective}. \s

In the following, we present results for the pair production of two
$H_1 \equiv h$ SM-like Higgs bosons, $gg \to hh$. The scan performed in the
NMSSM parameter space resulted in numerous scenarios, where the
heavier scalar Higgs bosons can decay into 
two on-shell Higgs bosons $hh$, leading to resonantly enhanced
contributions to the total pair production cross section. The program
{\tt HPAIR} computes the resonantly enhanced diagrams together with the
remaining diagrams up to NLO QCD by making use of Breit-Wigner propagators 
for the intermediate Higgs bosons. Alternatively, the non-resonant
contributions can be calculated by {\tt HPAIR} at NLO QCD while the resonant ones
are obtained by multiplying the cross section of a singly produced
heavy Higgs boson with its branching ratio into $hh$. For comparison, we have also
calculated the cross section in this approach, which of
course neglects off-shell effects and interference terms between
non-resonant and resonant contributions. The cross
section for single Higgs production is obtained with the
Fortran code {\tt HIGLU} \cite{Spira:1995mt} and is calculated at NLO
QCD to match the same order in $\alpha_s$ as in Higgs pair
production.\footnote{The program {\tt HIGLU} includes the QCD
  corrections to quark 
loops up to NNLO QCD, taking into account the full mass dependence at
NLO QCD \cite{nloqcdh}. The NLO QCD corrections to the squark loops
have been implemented in the limit of heavy squark masses
\cite{Dawson:1996xz}, which is a good approximation for squark masses
above about 400~GeV. For the finite squark mass effects on
  the NLO QCD corrections to Higgs production through gluon fusion, see
\cite{Muhlleitner:2006wx}.} In {\tt HIGLU} the Higgs Yukawa
couplings have been replaced by the corresponding couplings of the produced 
NMSSM Higgs boson. The non-resonant and
resonant contributions are then summed up to get the total cross
section. We have compared the results obtained within the two approaches,
and  the difference has been found to be less than 10\% for our scenarios. 
Note, that
the theoretical uncertainty on the NLO QCD double Higgs production
cross section due to unknown higher order corrections
\cite{Dawson:1998py} is larger than for single Higgs production at NLO
QCD \cite{Dittmaier:2011ti}. All results presented hereafter have been
obtained with {\tt HPAIR}. We emphasize that, if not stated otherwise,
in our calculation we have included the loop-corrected effective
trilinear Higgs couplings, presented in subsection 
\ref{sec:effective}, and the involved Higgs bosons both in the final
state and in the $s$-channel exchange are the loop-corrected Higgs
states. All involved loop quantities are taken at zero external momentum. 
\s

The hadronic cross sections for SM-like Higgs boson pair production
have been computed with MSTW 2008 NLO parton distribution functions (PDF) 
\cite{Martin:2009iq}. The factorization scale has been taken equal to
the renormalisation scale and set equal to the invariant mass of
the produced Higgs boson pair. In the code {\tt HPAIR}  top and bottom
quark pole masses are used as default inputs.

\subsection{Results for Higgs boson pair production}
In Table \ref{table_1} we show five sample scenarios compatible with
the constraints according to Eq.~(\ref{eq:cond}), where
$H_1$ is SM-like (points 1,2), 
$H_2$ is SM-like (points 4,5), and where $H_1$ and $H_2$ are close in mass
with $H_2$ being near 125~GeV (point 3). In all
scenarios the $H_3$ mass is large enough to allow for resonantly
enhanced $h$ pair production from the 
\begin{table}[h]
\begin{footnotesize}
\bc 
\vspace*{0.5cm}
\vspace*{0.5cm}
\begin{tabular}{|c|c|c|c|c|c|c|c|c|c|c|c|c|}
 \hline
{}
&$\tan\beta$
&$M_{H^\pm}$ 
&$\mu_{\eff}$
&$\la$
&$\kappa$ 
&$A_{\kappa}$
&$M_{\text{SUSY}}$
&$M_{H_1}$
&$M_{H_2}$
&$M_{H_3}$
&$M_{A_1}$
&$M_{A_2}$ \\
\hline \hline
Point 1 & 1.90 & 328  & 175 & 0.565 &0.406 &  -324  &710  & 124   & 146& 342 &324  &371   \\
Point 2 & 2.23 & 327  & 147  & 0.62 &0.32  &  -36.0  &738  & 125.7  & 143& 344 & 153  &333   \\ 
Point 3 & 2.28 & 278  & 121  & 0.507  & 0.349 & -116  &706 & 122   &125.8  & 294& 200 &280 \\ 
Point 4 & 2.69 &302   & 124  & 0.41 & 0.53 &  -533    &698   &  111  &
124.2  & 313 &  296&509 \\ 
Point 5 & 3.50 &310   & 113  & 0.23 & 0.53 &  -984    &686   &  110  & 124.4  & 311 &  300&871 \\ 
\hline 
\end{tabular}
\caption{\label{table_1}{List of parameter points for five
     sample scenarios. Particle masses and dimensionful parameters are
     given in GeV, and $m_t^{\scriptsize \mbox{eff}} = 155$~GeV,
     $m_b^{\scriptsize \mbox{eff}}= 2.47$~GeV.}} \ec 
\end{footnotesize}
\end{table} 
decay of $H_3 \to hh$, which can lead to a significant increase of the
Higgs pair production cross section depending on the trilinear Higgs
self-coupling $\lambda_{H_3 hh}$. Table \ref{table_2} contains the cross sections
including NLO QCD corrections for the production of a pair of SM-like
Higgs bosons $h$ in the five scenarios. Here $\sigma_T$ denotes the
cross sections calculated 
using the effective tree-level trilinear Higgs couplings $\lambda_{\scriptsize \mbox{NMSSM}}^{\scriptsize \mbox{eff}}
(\mbox{tree})$, while the cross section
values $\sigma_L$ use the effective loop-corrected trilinear Higgs
self-couplings $\lambda_{\scriptsize \mbox{NMSSM}}^{\scriptsize \mbox{eff}}
(\mbox{loop})$. The Higgs 
boson masses are taken at loop level in both cases. The quantity
$\delta$ is the difference between the two cross sections in terms of
$\sigma_T$,
\beq
\delta = \frac{\sigma_L - \sigma_T}{\sigma_T} \;, \quad
\mbox{with } \quad \sigma_L \equiv \sigma_{\lambda_{\scriptsize \mbox{NMSSM}}^{\scriptsize \mbox{eff}}
(\mbox{loop})} \quad\mbox{and}\quad \sigma_T \equiv
\sigma_{\lambda_{\scriptsize \mbox{NMSSM}}^{\scriptsize \mbox{eff}}
  (\mbox{tree})} \;.
\label{eq:deltasigma}
\eeq
The cross section values vary significantly in the various scenarios. 
As can be inferred from the table, the differences in the cross
sections due to the inclusion of loop corrections in the trilinear
Higgs self-couplings can be substantial, ranging from nearly 40\% to
almost 90\% in terms of the tree-level cross section for the chosen scenarios. \s

\begin{table}[h]
 \begin{footnotesize}
 \bc 
\begin{tabular}{|c|r|r|r|r|r|r|c|}
 \hline
{}
&$\si_\text{T} [fb]$ 
&$\si_\text{L} [fb]$
&$\de$\\
\hline \hline
Point 1    & 485.9(4) &  55.08(4) &  -0.89      \\
Point 2    & 462.9(3)  &254.2(2)&      -0.45  \\
Point 3   & 374.3(3)   &  175.5(1) & -0.53 \\ 
Point 4   &  99.30(7) &  28.36(2) & -0.71   \\
Point 5   & 17.52(1)  & 24.05(2)  & 0.37   \\ 

\hline 
\end{tabular}
\caption{\label{table_2}{The total cross sections in fb for $pp\to
     H_iH_i$ through gluon fusion 
     at $\sqrt{s}= 14\tev$, with $H_i$ being the SM-like Higgs
     boson,
     evaluated with tree-level ($\si_T$) and loop-corrected
     ($\si_L$) effective trilinear Higgs couplings. The
     deviation in the cross sections is quantified by
     $\de=(\si_L-\si_T)/\si_T$.}}
\ec
 \end{footnotesize}
\end{table} 

In Fig.~\ref{fig:hpair} (left) we show the gluon fusion production
cross section of a pair of SM-like Higgs bosons $h$ calculated with
the effective loop-corrected trilinear Higgs couplings, as a function of the ratio $\lambda_{\scriptsize \mbox{NMSSM}}^{\scriptsize \mbox{eff}} (\mbox{loop})/\lambda_{\scriptsize
  \mbox{SM}}^{\scriptsize \mbox{eff}} (\mbox{loop})$. Note that
in the ratio $\lambda_{\scriptsize\mbox{NMSSM}}$ refers to the trilinear
Higgs self-coupling of the SM-like Higgs boson $h$. In the case 3 of
two Higgs bosons nearby 125~GeV, the name SM-like Higgs boson refers
to the Higgs boson with the most SM-like Higgs Yukawa coupling to top
quarks, as this coupling determines the dominant Higgs production
cross section through gluon fusion and has the major impact on the
Higgs production rate. The c.m. energy
has been taken to be 14~TeV. Shown are cross sections for scenarios
with $H_1 \equiv h$ (case 1), with  $H_2 \equiv h$ (case2) or with two Higgs bosons
with mass around 125~GeV building up the Higgs signal (case
3). The two black stars refer to scen1 and scen2 defined in
section~\ref{sec:resdecays}. Also plotted as a horizontal line is the SM
cross section evaluated 
with $\lambda_{\scriptsize \mbox{SM}}^{\scriptsize \mbox{eff}}
(\mbox{loop})$. 
The cross section amounts to 35~fb compared to 34~fb evaluated with the tree-level
coupling \cite{Baglio:2012np}. Furthermore, we plot as horizontal
lines the results for the SM cross section in
case of a variation of the effective loop-corrected trilinear coupling
in terms of $\lambda_{\scriptsize \mbox{SM}}^{\scriptsize \mbox{eff}} (\mbox{loop})$
by a factor 2 and by a factor -1 and in case of $\lambda_{\scriptsize
  \mbox{SM}}^{\scriptsize \mbox{eff}} (\mbox{loop})= 0$. 
They show the sensitivity of the SM cross section to the
trilinear Higgs self-coupling. In other words, they give an idea of
how precisely the cross section has to be measured to achieve a certain
accuracy in the trilinear SM Higgs self-coupling extracted from it. Many 
of the NMSSM cross section values lie nearby the SM result,
independently of the ratio between the NMSSM and SM effective
loop-corrected trilinear couplings, thus mimicking the SM case. However, there are also
parameter points where the NMSSM and SM cross section differ
significantly, so that a distinction of the two models would be
possible. Loop corrections play here a crucial role: 
\begin{figure}[h]
  \centering
\includegraphics[width=0.45\textwidth]{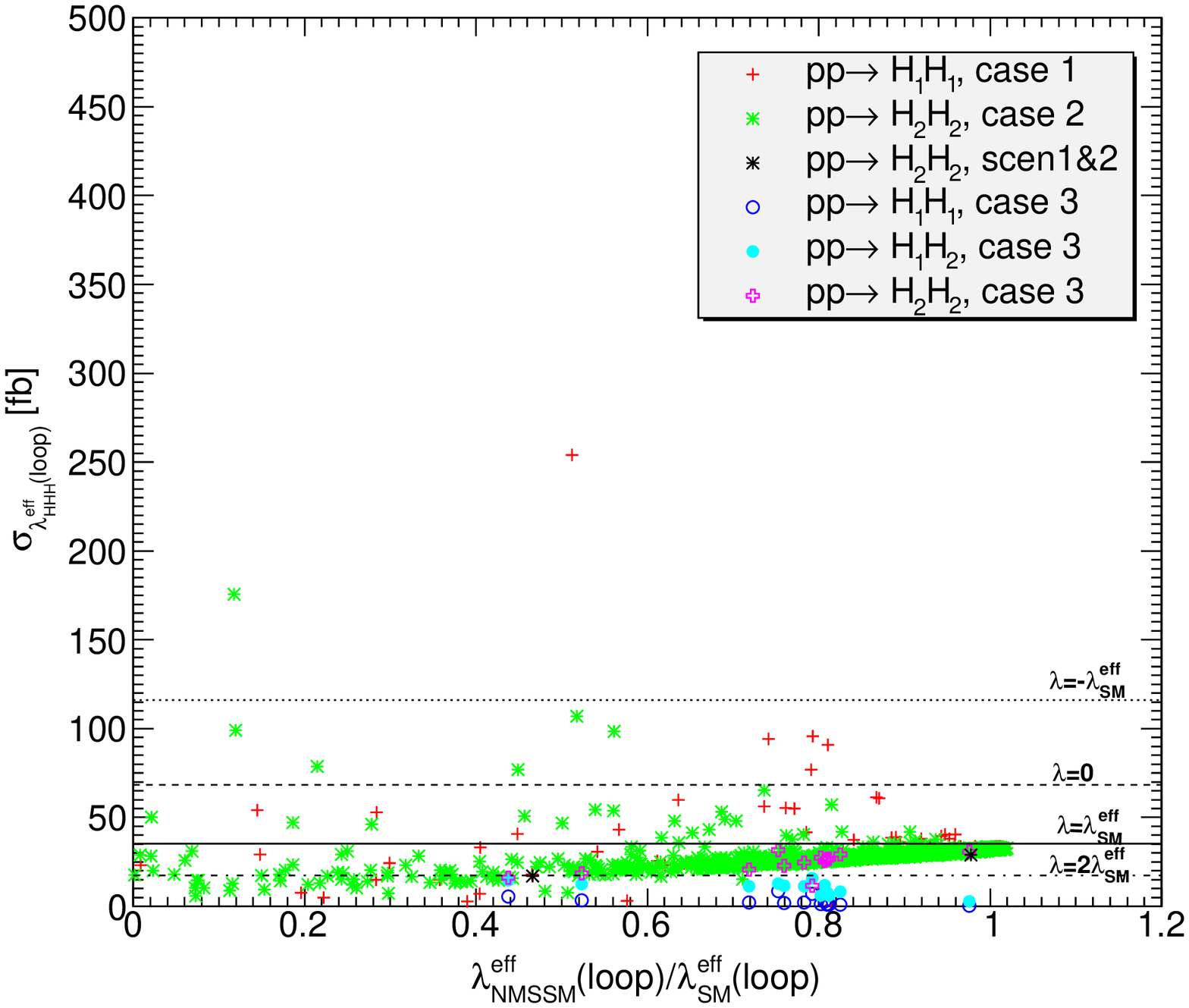}\label{fig:hpair1}\quad
\includegraphics[width=0.45\textwidth]{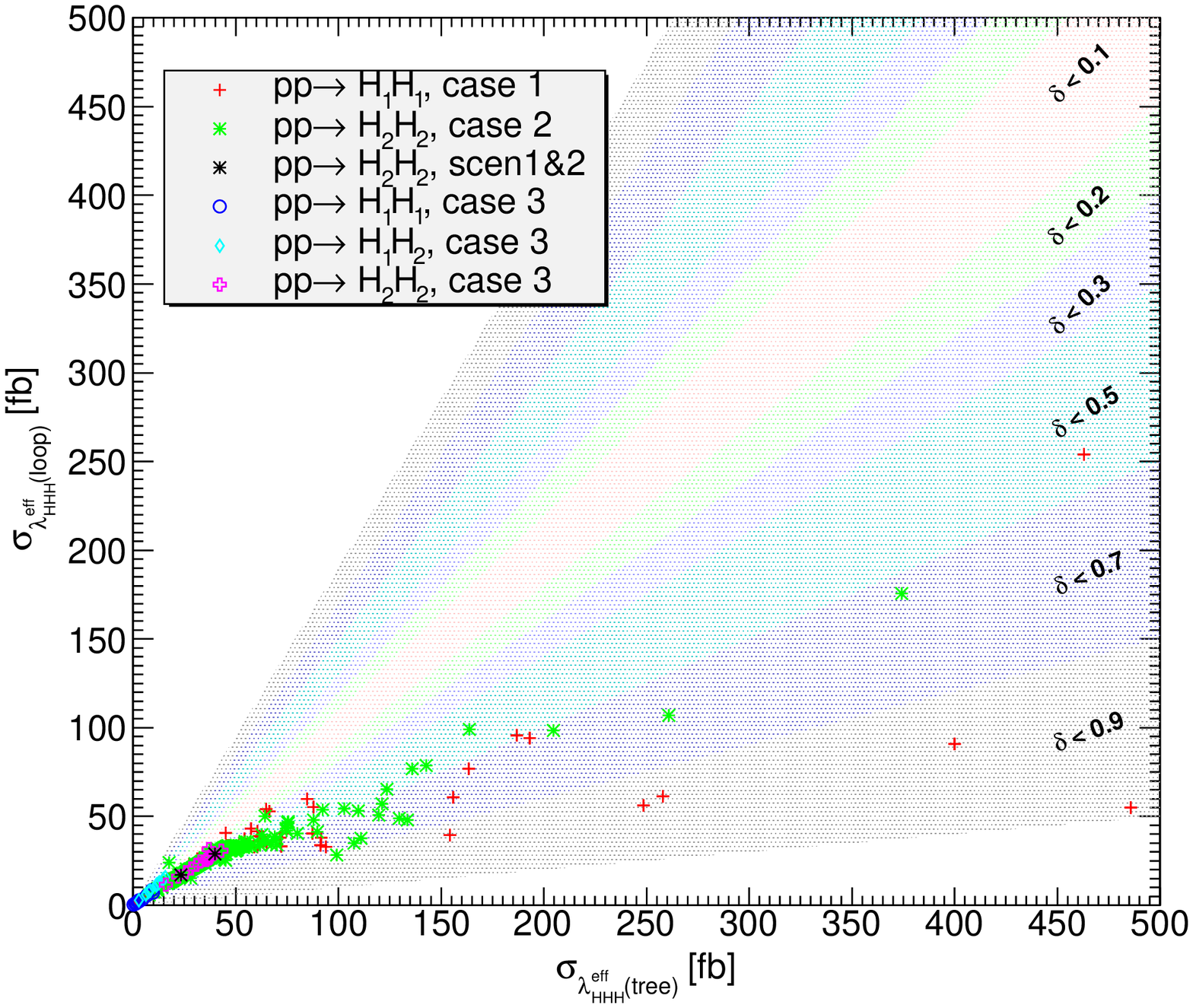}\label{fig:hpair2}
  \caption{Cross section of SM-like double Higgs production through
    gluon fusion evaluated with the
    effective loop-corrected trilinear Higgs self-couplings as a
    function of $\lambda_{\scriptsize \mbox{NMSSM}}^{\scriptsize \mbox{eff}}
(\mbox{loop})/\lambda_{\scriptsize \mbox{SM}}^{\scriptsize \mbox{eff}}
(\mbox{loop})$ (left) and plotted against the corresponding cross
    section evaluated with the tree-level effective trilinear Higgs
    self-couplings (right). Shown are scenarios with
    $h\equiv H_1$ (case 1), $h\equiv H_2$ (case 2, scen1, scen2), $H_1$ and
    $H_2$ close in mass near 125 GeV (case 3). Left: Also shown is the SM
    cross section value evaluated with the effective loop-corrected
    trilinear Higgs self-coupling (full line) and for Higgs
    self-coupling variations in terms of $\lambda_{\scriptsize
      \mbox{SM}}^{\scriptsize \mbox{eff}} (\mbox{loop})$ by a factor 2
    (dot-dashed), a factor -1 (dotted) and for vanishing coupling
    (dashed). Right: $\delta \equiv (\sigma_L-\sigma_T)/\sigma_T$.}
 \label{fig:hpair}
\end{figure}
In Fig.~\ref{fig:hpair} (right) we show for the cases 1-3 the SM-like Higgs pair
production cross section of the NMSSM evaluated with the effective
loop-corrected trilinear Higgs self-couplings versus the cross section
evaluated with the effective tree-level couplings. The difference
between the cross sections is again quantified by $\delta$, defined
in~Eq.~(\ref{eq:deltasigma}). 
There are many scenarios with $\delta$ up to 50\% and even
higher. These large deviations can be traced back to the large
deviations between the tree-level and loop-corrected branching ratios
of the $H_3$ decay into two SM-like Higgs bosons $h$, {\it
  cf.}~Fig.~\ref{fig:HTO2HSM}.\footnote{In the narrow-width
  approximation the Higgs pair production cross section is given by
  single Higgs production times the branching ratio BR($H_3 \to hh$).}
The fact that, depending on whether or not the loop corrections to the
trilinear Higgs self-couplings are included the scenario is
distinguishable or not from the SM, clearly shows the importance of the loop corrections
to obtain predictions useful for the experiments.

\section{Conclusions \label{sec:conclusions}}
The discovery of a new scalar particle by the LHC experiments ATLAS
and CMS has triggered a lot of activities to determine the properties
of this particle. While the analyses of the so far accumulated data
strongly suggest that it is indeed the Higgs boson, {\it i.e.}~the
particle responsible for the creation of particle masses without
violating gauge principles, the accumulation of more data is necessary for
its interpretation with respect to other models than the SM. The question has to
be clarified if it is the Higgs boson of the SM or of some extensions
beyond the SM. Among these, supersymmetric theories represent one of
the most intensely studied model classes. The Higgs sector of the NMSSM
consists of seven Higgs bosons entailing a rich phenomenology with
possible Higgs-to-Higgs decays and resonantly enhanced double Higgs
production cross sections due to heavier Higgs bosons decaying into a
pair of light Higgs particles. \s

In order to properly interpret the
experimental data and distinguish between different models the precise
prediction of the Higgs parameters such as masses and couplings,
including higher order corrections, is indispensable. The Higgs boson
masses and Higgs self-couplings are related to each other via the
Higgs potential. The prediction of loop-corrected Higgs boson masses
necessitates also the inclusion of loop corrections to the Higgs
self-couplings for a consistent analysis of the Higgs data. Having
calculated in previous works the one-loop corrected NMSSM Higgs boson
masses, in this contribution we extend our program of the calculation
of loop-corrected NMSSM Higgs parameters to the computation of
loop-corrected trilinear Higgs self-couplings. \s

The inclusion of loop corrections turns out to be important. We found
for example that non-SM Higgs decays of the 
125 GeV NMSSM Higgs boson into a pair of lighter Higgs bosons 
could be large enough to decrease the
signal strengths into SM particle final states so that they are not
compatible any more with the experimental results. The inclusion
of loop corrections to the trilinear couplings can, however, reduce
these rates to a level where the theoretical predictions are in
accordance with the experimental findings. In principle this could
also work the other way around, {\it i.e.}~a scenario could be allowed
if the tree-level trilinear Higgs couplings are used, but excluded
in case of loop-corrected couplings. Trilinear Higgs self-couplings
of course also play a role in the decays of heavier non-SM-like Higgs
bosons into light Higgs pairs and their possible detection via these
decays. \s

Trilinear Higgs self-couplings enter the production of Higgs boson
pairs, so that they can be 
extracted from the measurement of these processes. Once the (trilinear
and quartic) Higgs 
self-couplings are known, the Higgs potential can be reconstructed to
perform the ultimate step in the experimental verification of the Higgs
mechanism. We found that the inclusion of loop corrections to the
trilinear Higgs self-couplings can alter the Higgs pair production
cross sections through gluon fusion substantially. In many of the
scenarios passing the constraints, the production
cross section of a pair of SM-like Higgs bosons with mass of 125 GeV
can be different enough to distinguish it from SM Higgs boson pair
production. This depends of course also on the experimental accuracy which
can be reached in these processes and which relies on analyses to be
performed at the high-energy and high-luminosity run of the LHC.\s

In summary, the computation of loop corrections to Higgs
boson self-couplings and their inclusion in the analyses of the
experimental results is crucial, in particular for the proper
interpretation of the
Higgs data with respect to the exclusion or non-exclusion of NMSSM
parameter scenarios and/or with respect to the correct interpretation of Higgs
pair production processes.

\subsubsection*{Acknowledgments}
DTN, MMM and KW are supported by the DFG SFB/TR9  ``Computational
Particle Physics''. We thank Julien Baglio and Ramona Gr\"ober for discussions.
We would like to thank Pietro Slavich for the communication concerning
the two-loop contribution to the NMSSM Higgs boson masses.


\begin{appendix}
\section{Tree-level trilinear Higgs couplings \label{app:1}}
In this appendix we present the trilinear Higgs self-couplings of the
interaction eigenstates of the NMSSM Higgs sector. The tree-level
trilinear couplings of the scalar mass eigenstates $h_i$ ($i=1,2,3$)
are obtained by applying the tree-level rotation matrices ${\cal
  R}^S$, Eq.~(\ref{eq:rotationS}), on the three interaction eigenstates, 
\beq
\lambda_{h_i h_j h_k} = {\cal R}_{ii'}^S {\cal R}_{jj'}^S {\cal
  R}_{kk'}^S \lambda_{i'j'k'}^{hhh} \;, \qquad i,j,k=1,2,3 \; , \; i',j',k'=1,2,3\;.
\eeq
The indices $i',j',k'$ refer to the interaction eigenstates, and we
have the following correspondences $1\mathrel{\widehat{=}} h_d$,
$2\mathrel{\widehat{=}} h_u$, $3\mathrel{\widehat{=}} h_s$. The
couplings $\lambda_{i'j'k'}^{hhh}$ are symmetric in the three
indices. Using the short-hand notations $c_\beta\equiv \cos\beta, \,
s_\beta\equiv\sin\beta,\, t_\beta \equiv \tan\beta$, we have
\begin{align} 
\la_{111}^{hhh} &= \fr{3 c_\beta M_Z^2 }{v},&\la_{112}^{hhh} &=\braket{-\fr{M_Z^2}{v} + \lambda^2  v}s_\beta,
&\la_{113}^{hhh}&=\sqrt{2} \lambda \mu_\eff,\crn
 \la_{122}^{hhh}&=\braket{-\fr{M_Z^2}{v} + \lambda^2  v}c_\beta,&
\la_{123}^{hhh} &= -\fr{A_\lambda \lambda}{\sqrt{2}} - \sqrt{2} \kappa \mu_\eff,&
\la_{133}^{hhh} &=(c_\beta \lambda  - \kappa  s_\beta) v\lambda,\crn 
\la_{222}^{hhh} &=\fr{3 M_Z^2 s_\beta} {v},& \la_{223}^{hhh}&=
\sqrt{2} \lambda \mu_\eff, &
\la_{233}^{hhh} &=(-c_\beta \kappa   + \lambda s_\beta) v\lambda,\crn
\la_{333}^{hhh}&= \sqrt{2}\kappa\braket{\fr{6  \kappa \mu_\eff}\lambda + \sqrt{2} A_\kappa }.
\end{align}
The trilinear couplings of one CP-even Higgs boson with two CP-odd
Higgs states $a_l$ ($l=1,2$) are obtained from the interaction eigenstates through 
\beq
\lambda_{h_i a_l a_m} = {\cal R}_{ii'}^S {\cal R}_{ll'}^P {\cal
  R}_{mm'}^P \lambda^{haa}_{i'l'm'} \; , \qquad i,i',l',m'=1,2,3 \; ,
\quad l,m=1,2 \;.
\eeq
Here, for the indices $i',l',m'$ we have the correspondences $1
\mathrel{\widehat{=}} a$, $2 \mathrel{\widehat{=}} a_s$, $3
\mathrel{\widehat{=}} G$. The couplings $\lambda^{haa}_{i'l'm'}$ are
symmetric with respect to an exchange of the last two indices
$l',m'$, and
\begin{align}
\la_{111}^{haa}&=-\fr{c_\beta c_{2\beta} M_Z^2}{v}  + c_\beta^3 \lambda^2 v, &\la_{112}^{haa}&= c_\beta \braket{\fr{ A_\lambda \lambda }{\sqrt{2}} - \sqrt{2} \kappa \mu_\eff},\crn
\la_{113}^{haa} &=c_\beta^2 s_\beta\braket{\fr{2  M_Z^2 }{v} -  \lambda^2  v},&\la_{122}^{haa}&=(c_\beta \lambda  + \kappa  s_\beta) \lambda v,\crn
\la_{123}^{haa}&= s_\beta\braket{-\fr{ A_\lambda \lambda }{\sqrt{2}} + \sqrt{2} \kappa \mu_\eff}, &\la_{133}^{haa}&=\fr{c_\beta c_{2\beta} M_Z^2} {v} + c_\beta s_\beta^2 \lambda^2  v,\crn
\la_{211}^{haa}&=\fr{s_\beta c_{2\beta} M_Z^2} {v} +  \lambda^2 s_\beta^3 v, &\la_{212}^{haa}&= s_\beta\braket{\fr{ A_\lambda \lambda }{\sqrt{2}} - \sqrt{2} \kappa \mu_\eff},\crn
\la_{213}^{haa}&=-c_\beta s^2_\beta\braket{\fr{2  M_Z^2 }{v} -  \lambda^2  v},&\la_{222}^{haa}&=(c_\beta \kappa +s_\beta\la )\la v,\crn
\la_{223}^{haa}&= c_\beta\braket{\fr{ A_\lambda \lambda }{\sqrt{2}} -
  \sqrt{2} \kappa \mu_\eff}, & 
\la_{233}^{haa}&=-\fr{c_{2\beta}s_\beta M_Z^2 }{v}  + c_\beta^2
s_\beta \lambda^2  v,
\crn
\la_{311}^{haa}&=  s_{2\beta}\braket{ \fr{ A_\lambda  \lambda}{ \sqrt{2}}  +
   \sqrt{2} \kappa \mu_\eff}  + \sqrt{2} \lambda \mu_\eff, &\la_{312}^{haa}&= -\kappa \lambda v,\crn
\la_{313}^{haa}&=c_{2\beta}\braket{\fr{A_\lambda  \lambda}{\sqrt{2}} + \sqrt{2}  \kappa \mu_\eff}, &\la_{322}^{haa}&=\fr{2 \sqrt2 \kappa^2 \mu_\eff}{\lambda} - \sqrt2 \kappa A_\kappa,\crn
\la_{323}^{haa}&=0, &\la_{333}^{haa}&= - s_{2\beta} \braket{ \fr{ A_\lambda  \lambda}{ \sqrt{2}}  +
   \sqrt{2} \kappa \mu_\eff}  + \sqrt{2} \lambda \mu_\eff.
\end{align}
\end{appendix}


\end{document}